\documentstyle[12pt,epsfig]{article}

\newcommand{\agt}{\,\rlap{\lower 3.5 pt \hbox{$\mathchar \sim$}} \raise 1pt
 \hbox {$>$}\,}
\newcommand{\alt}{\,\rlap{\lower 3.5 pt \hbox{$\mathchar \sim$}} \raise 1pt
 \hbox {$<$}\,}

\catcode`@=11
\newcount\@tempcntc
\def\@citex[#1]#2{\if@filesw\immediate\write\@auxout{\string\citation{#2}}\fi
  \@tempcnta\z@\@tempcntb\m@ne\def\@citea{}\@cite{\@for\@citeb:=#2\do
    {\@ifundefined
       {b@\@citeb}{\@citeo\@tempcntb\m@ne\@citea\def\@citea{,}{\bf ?}\@warning
       {Citation `\@citeb' on page \thepage \space undefined}}%
    {\setbox\z@\hbox{\global\@tempcntc0\csname b@\@citeb\endcsname\relax}%
     \ifnum\@tempcntc=\z@ \@citeo\@tempcntb\m@ne
       \@citea\def\@citea{,}\hbox{\csname b@\@citeb\endcsname}%
     \else
      \advance\@tempcntb\@ne
      \ifnum\@tempcntb=\@tempcntc
      \else\advance\@tempcntb\m@ne\@citeo
      \@tempcnta\@tempcntc\@tempcntb\@tempcntc\fi\fi}}\@citeo}{#1}}
\def\@citeo{\ifnum\@tempcnta>\@tempcntb\else\@citea\def\@citea{,}%
  \ifnum\@tempcnta=\@tempcntb\the\@tempcnta\else
   {\advance\@tempcnta\@ne\ifnum\@tempcnta=\@tempcntb \else \def\@citea{--}\fi
    \advance\@tempcnta\m@ne\the\@tempcnta\@citea\the\@tempcntb}\fi\fi}
\catcode`@=12

\begin{document}

\title{
\vskip-3cm{\baselineskip14pt
\centerline{\normalsize DESY 03-107\hfill ISSN 0418-9833}
\centerline{\normalsize hep-ph/0308164\hfill}
\centerline{\normalsize August 2003\hfill}
}
\vskip1.5cm
Prompt $J/\psi$ production in charged-current deep-inelastic scattering}
\author{
{\sc Bernd A. Kniehl, Lennart Zwirner}\\
{\normalsize II. Institut f\"ur Theoretische Physik, Universit\"at Hamburg,}\\
{\normalsize Luruper Chaussee 149, 22761 Hamburg, Germany}}

\date{}

\maketitle

\thispagestyle{empty}

\begin{abstract}
We calculate the cross section of $J/\psi$ inclusive production in
lepton-nucleon deep-inelastic scattering via the weak charged current within
the factorization formalism of nonrelativistic quantum chromodynamics.
The hadronic system produced in association with the $J/\psi$ meson can be of
light-quark origin or contain charm.
We take into account both direct production and feed-down from
directly-produced heavier charmonia.
We provide the cross sections of all contributing partonic subprocesses in
analytic form.
We present theoretical predictions for the $J/\psi$ transverse-momentum and
rapidity distributions, which can be measured in current neutrino-nucleon
scattering experiments at CERN and Fermilab, and possibly at the future
electron-proton collider THERA at DESY.
We find the cross section to be at the edge of observability at DESY HERA.

\medskip

\end{abstract}

\newpage

\section{Introduction}
\label{sec:one}

Since the discovery of the $J/\psi$ meson in 1974, charmonium has provided a
useful laboratory for quantitative tests of quantum chromodynamics (QCD) and,
in particular, of the interplay of perturbative and nonperturbative phenomena.
The factorization formalism of nonrelativistic QCD (NRQCD) \cite{bbl} provides
a rigorous theoretical framework for the description of heavy-quarkonium
production and decay.
This formalism implies a separation of short-distance coefficients, which can 
be calculated perturbatively as expansions in the strong-coupling constant
$\alpha_s$, from long-distance matrix elements (MEs), which must be extracted
from experiment.
The relative importance of the latter can be estimated by means of velocity
scaling rules, i.e., the MEs are predicted to scale with a definite power of
the heavy-quark ($Q$) velocity $v$ in the limit $v\ll1$.
In this way, the theoretical predictions are organized as double expansions in
$\alpha_s$ and $v$.
A crucial feature of this formalism is that it takes into account the complete
structure of the $Q\overline{Q}$ Fock space, which is spanned by the states
$n={}^{2S+1}L_J^{(\zeta)}$ with definite spin $S$, orbital angular momentum
$L$, total angular momentum $J$, and colour multiplicity $\zeta=1,8$.
In particular, this formalism predicts the existence of colour-octet (CO)
processes in nature.
This means that $Q\overline{Q}$ pairs are produced at short distances in
CO states and subsequently evolve into physical, colour-singlet (CS) quarkonia
by the nonperturbative emission of soft gluons.
In the limit $v\to0$, the traditional CS model (CSM) \cite{ber} is recovered.
The greatest triumph of this formalism was that it was able to correctly 
describe \cite{bra} the cross section of inclusive charmonium
hadroproduction measured in $p\overline{p}$ collisions at the Fermilab
Tevatron \cite{abe}, which had turned out to be more than one order of
magnitude in excess of the theoretical prediction based on the CSM.

Apart from this phenomenological drawback, the CSM also suffers from severe
conceptual problems indicating that it is incomplete.
These include the presence of logarithmic infrared divergences in the
${\cal O}(\alpha_s)$ corrections to $P$-wave decays to light hadrons and in
the relativistic corrections to $S$-wave annihilation \cite{bar}, and the lack
of a general argument for its validity in higher orders of perturbation
theory.
While the $k_T$-factorization \cite{sri} and hard-comover-scattering
\cite{hoy} approaches manage to bring the CSM prediction much closer to the
Tevatron data, they do not cure the conceptual defects of the CSM.
The colour evaporation model \cite{cem}, which is intuitive and useful for
qualitative studies, also leads to a significantly better description of the
Tevatron data, but it is not meant to represent a rigorous framework for
perturbation theory.
In this sense, a coequal alternative to the NRQCD factorization formalism is 
presently not available.

In order to convincingly establish the phenomenological significance of the
CO processes, it is indispensable to identify them in other kinds of
high-energy experiments as well.
Studies of charmonium production in $ep$ photoproduction, $ep$ and $\nu N$
deep-inelastic scattering (DIS), $e^+e^-$ annihilation, $\gamma\gamma$
collisions, and $b$-hadron decays may be found in the literature; see
Ref.~\cite{fle} and references cited therein.
Furthermore, the polarization of $\psi^\prime$ mesons produced directly and of
$J/\psi$ mesons produced promptly, i.e., either directly or via the feed-down
from heavier charmonia, which also provides a sensitive probe of CO processes,
was investigated \cite{ben,van,bkl,lee}.
Until very recently, none of these studies was able to prove or disprove the
NRQCD factorization hypothesis.
However, H1 data of $e+p\to e+J/\psi+X$ in DIS at HERA \cite{h1} and DELPHI
data of $\gamma+\gamma\to J/\psi+X$ at LEP2 \cite{delphi} provide first
independent evidence for it \cite{ep,gg}.

Recently, we studied $J/\psi$ inclusive production in $ep$ DIS via the
electromagnetic current \cite{ep} and in $\nu N$ DIS via the weak neutral
current (NC) \cite{nun}.
In both cases, we allowed for an additional hadron jet $j$ in the final state,
i.e., we considered the processes $e+p\to e+J/\psi+j+X$ and
$\nu+N\to\nu+J/\psi+j+X$, respectively, where $X$ denotes the nucleus remnant.
In this way, the inelasticity variable $z$, which measures the fraction of
energy transferred from the virtual photon ($\gamma^\star$) or $Z$ boson
($Z^\star$) to the $J/\psi$ meson in the nucleus rest frame, can take values
below unity, away from the endpoint $z=1$, where diffractive production takes
place.
At the same time, the $J/\psi$ meson can acquire finite transverse momentum
$p_T^\star$ in the $\gamma^\star p$ and $Z^\star N$ centre-of-mass (CM)
frames, and the hadronic system $X^\prime$ consisting of $j$ and $X$ can
acquire finite mass $M_{X^\prime}$.
By the same token, diffractive events can be eliminated from the experimental
data sample by applying appropriate acceptance cuts on $z$, $p_T^\star$, or
$M_{X^\prime}$.

In this paper, we complete our study of $J/\psi$ inclusive production in DIS
by considering the processes $e+p\to\nu+J/\psi+j+X$ and
$\nu+N\to e+J/\psi+j+X$, which are mediated by the weak charged current (CC).
The former process can, in principle, be analyzed experimentally at the
hadron-electron ring accelerator (HERA), with CM energy $\sqrt S=318$~GeV, in
operation at DESY, and at THERA \cite{thera}, a future hadron-electron
supercollider with $\sqrt S=1$~TeV that uses the HERA proton beam and the
electron beam of the TeV-Energy Superconducting Linear Accelerator (TESLA),
which is presently being designed and planned at DESY.
The latter process can be investigated by the NuTeV Collaboration at Femilab
(Experiment E815) \cite{nutev} and by the CHORUS \cite{chorus} and NOMAD
\cite{nomad} Collaborations at CERN.
Since these experiments utilize $\nu_\mu$ or $\overline{\nu}_\mu$ beams and
$J/\psi$ mesons are preferably identified through their decays to $e^+e^-$ or
$\mu^+\mu^-$ pairs, signal events would contain spectacular $\mu^\pm e^+e^-$
or $\mu^\pm\mu^+\mu^-$ trileptons together with hadronic debris from the
target.
In fact, events of both the former \cite{bcs} and latter \cite{tri} types were
observed in $\nu N$ fixed-target experiments at CERN and Fermilab.
A theoretical study of the processes $\nu_\mu+N\to\mu^-+J/\psi+X$ and
$\overline{\nu}_\mu+N\to\mu^++J/\psi+X$ may be found in Ref.~\cite{god}, where
the $J/\psi$ mesons are assumed to arise electronically, i.e., from virtual
photons that are emitted from charged leptons, quarks, or $W$ bosons involved
in the hard scatterings.
To our knowledge, hadronic production mechanisms in the framework of NRQCD or
the CSM have not yet been considered in the literature.
On the other hand, open $c\overline{c}$ production in $\nu N$ CC DIS was
studied in QCD \cite{gol}.

Diffractive contributions to prompt $J/\psi$ production in CC DIS should be
greatly suppressed.
In contrast to the photon and $Z$ boson, the $W$ boson, being charged, cannot
directly fluctuate into charmonium.
A secondary CC interaction is necessary for that.
We refer the interested reader to the pioneering literature \cite{ble}.
An alternative set of Feynman diagrams that allows charmonium to diffractively
scatter off the hadron target involves a timelike $W^\star$ boson, which
decays.
The charmonium state would then originate from a $\gamma^\star$ or $Z^\star$
boson that is coupled to the lepton line, the $W^\star$ boson, or the decay
products of the latter.
Both sets of Feynman diagrams interfere, and the resulting cross section is
suppressed by two powers of Sommerfeld's fine-structure constant $\alpha$
relative to its counterparts in electromagnetic or NC DIS.
For this reason, diffractive processes are disregarded in the following.
In particular, there is no need for acceptance cuts on $z$, $p_T^\star$, or
$M_{X^\prime}$.

The $J/\psi$ mesons can be produced directly; or via radiative or hadronic
decays of heavier charmonia, such as $\chi_{cJ}$ and $\psi^\prime$ mesons; or
via weak decays of $b$ hadrons.
The respective decay branching fractions are
$B(\chi_{c0}\to J/\psi+\gamma)=(1.02\pm0.17)\%$,
$B(\chi_{c1}\to J/\psi+\gamma)=(31.6\pm3.2)\%$,
$B(\chi_{c2}\to J/\psi+\gamma)=(18.7\pm2.0)\%$,
$B(\psi^\prime\to J/\psi+X)=(55.7\pm2.6)\%$, and
$B(B\to J/\psi+X)=(1.15\pm0.06)\%$ \cite{pdg}.
The $b$ hadrons can be detected by looking for displaced decay vertices with
dedicated vertex detectors.
Therefore and because of the smallness of $B(B\to J/\psi+X)$, $J/\psi$
production through $b$-hadron decay is not considered here.
The cross sections of the four residual indirect production channels may be
approximated by multiplying the direct-production cross sections of the
respective intermediate charmonia with their decay branching fractions to
$J/\psi$ mesons.

This paper is organized as follows.
In Section~\ref{sec:ana}, we list our formulas for the cross sections of
prompt $J/\psi$ production in CC DIS.
Lengthy expressions are relegated to the Appendix.
In Section~\ref{sec:num}, we present our numerical results and discuss their
phenomenological implications.
Our conclusions are summarized in Section~\ref{sec:con}.

\section{Analytic results}
\label{sec:ana}

In this section, we present our analytic results for the cross sections of
\begin{eqnarray}
\nu_l+N&\to&l^-+H+j+X,
\label{eq:N}\\
\overline{\nu}_l+N&\to&l^++H+j+X,
\label{eq:AN}\\
l^-+N&\to&\nu_l+H+j+X,
\label{eq:L}\\
l^++N&\to&\overline{\nu}_l+H+j+X,
\label{eq:AL}
\end{eqnarray}
in CC DIS, where $l=e,\mu,\tau$, $N$ is the hadron target, and 
$H=J/\psi,\chi_{cJ},\psi^\prime$ with $J=0,1,2$.

We work in the fixed-flavour-number scheme, i.e., we have $n_f=3$ active
quark flavours $q=u,d,s$ in the nucleus $N$.
We treat the nucleus $N$ and the quarks $q$ as massless.
The charm quark $c$ and antiquark $\overline{c}$, with mass $m_c$, only appear
in the final state.
The $c\overline{c}$ Fock states contributing at lowest order (LO) in $v$ are
$n={}^3\!S_1^{(1)},{}^1\!S_0^{(8)},{}^3\!S_1^{(8)},{}^3\!P_J^{(8)}$ if
$H=J/\psi,\psi^\prime$ and
$n={}^3\!P_J^{(1)},{}^3\!S_1^{(8)}$ if $H=\chi_{cJ}$.
Their MEs satisfy the multiplicity relations
\begin{eqnarray}
\left\langle{\cal O}^\psi\left[{}^3\!P_J^{(8)}\right]\right\rangle
&=&(2J+1)
\left\langle{\cal O}^\psi\left[{}^3\!P_0^{(8)}\right]\right\rangle,
\nonumber\\
\left\langle{\cal O}^{\chi_{cJ}}\left[{}^3\!P_J^{(1)}\right]\right\rangle
&=&(2J+1)
\left\langle{\cal O}^{\chi_{c0}}\left[{}^3\!P_0^{(1)}\right]\right\rangle,
\nonumber\\
\left\langle{\cal O}^{\chi_{cJ}}\left[{}^3\!S_1^{(8)}\right]\right\rangle
&=&(2J+1)
\left\langle{\cal O}^{\chi_{c0}}\left[{}^3\!S_1^{(8)}\right]\right\rangle,
\label{eq:mul}
\end{eqnarray}
which follow to LO in $v$ from heavy-quark spin symmetry.

We are thus led to consider the following partonic subprocesses
\begin{eqnarray}
\nu_l+a&\to&l^-+c\overline{c}[n]+b,
\label{eq:n}\\
\overline{\nu}_l+\overline{a}&\to&l^++c\overline{c}[n]+\overline{b},
\label{eq:an}\\
l^-+\overline{a}&\to&\nu_l+c\overline{c}[n]+\overline{b},
\label{eq:l}\\
l^++a&\to&\overline{\nu}_l+c\overline{c}[n]+b,
\label{eq:al}
\end{eqnarray}
where $(a,b)=(d,u),(\overline{u},\overline{d}),(s,c)$ for the transitions
diagonal w.r.t.\ the Cabibbo-Kobayashi-Maskawa (CKM) flavour mixing and
$(a,b)=(s,u),(\overline{u},\overline{s}),(d,c)$ for the CKM-suppressed ones.
Here, it is understood that $\overline{\overline{q}}=q$.
Typical Feynman diagrams are depicted in Figs.~\ref{fig:fey} (a) and (b).
If $b=c$ [see Fig.~\ref{fig:fey} (b)], then the jet $j$ contains a charmed
hadron, which can be identified experimentally.
In this case, all the $c\overline{c}$ Fock states enumerated above contribute.
In particular, this production mode is also possible within the CSM.
By contrast, if $b=u,\overline{d},\overline{s}$ [see Fig.~\ref{fig:fey} (a)],
then only the CO channel $n={}^3\!S_1^{(8)}$ contributes, i.e., the prompt
production of a $J/\psi$ meson in association with an otherwise charmless
hadronic system in CC DIS is a pure CO process.

We now define the kinematics, considering process (\ref{eq:N}) for
definiteness.
As in Refs.~\cite{ep,nun}, we denote the four-momenta of $\nu_l$, $N$, $l^-$,
$H$, and $j$ by $k$, $P$, $k^\prime$, $p_H$, and $p^\prime$, respectively.
The parton struck by the virtual $W$ boson ($W^\star$) carries four-momentum
$p=xP$.
We take the mass $M$ of $H$ to be $M=2m_c$.
The invariant mass $m$ of $j$ is $m=m_c$ if $b=c$ and zero otherwise.
In our approximation, $X$ has zero invariant mass, $M_X^2=(P-p)^2=0$.
The CM energy square of the $\nu_lN$ collision is $S=(k+P)^2$.
The four-momentum of $W^\star$ is $q=k-k^\prime$.
As usual, we define $Q^2=-q^2>0$, $y=q\cdot P/k\cdot P$, and the inelasticity
variable $z=p_H\cdot P/q\cdot P$.
In the rest frame of $N$, $y$ and $z$ measure the relative lepton energy loss
and the fraction of the $W^\star$ energy transferred to $H$, respectively.
The $W^\star N$ CM energy square is $W^2=(q+P)^2=yS-Q^2$.
The system $X^\prime$ consisting of $j$ and $X$ has invariant mass square
$M_{X^\prime}^2=(q+P-p_H)^2=m^2+(1-x)y(1-z)S$.
As usual, we define the partonic Mandelstam variables as
$\hat s=(q+p)^2=xyS-Q^2$, $\hat t=(q-p_H)^2=m^2-xy(1-z)S$, and
$\hat u=(p-p_H)^2=M^2-xyzS$.
By four-momentum conservation, we have $\hat s+\hat t+\hat u=M^2+m^2-Q^2$.
In the $W^\star N$ CM frame, $H$ has transverse momentum and
rapidity
\begin{eqnarray}
p_T^\star&=&\frac{\sqrt{\hat s\left(\hat t\hat u-M^2m^2\right)
+Q^2\left(\hat tM^2+\hat um^2-2M^2m^2\right)}}{\hat s+Q^2},
\label{eq:ptcms}\\
y_H^\star&=&\frac{1}{2}\ln\frac{\hat s\left(M^2-\hat u\right)}
{\hat s\left(M^2-\hat t\right)+Q^2(M^2-m^2)}+\frac{1}{2}\ln\frac{W^2}{\hat s},
\label{eq:ycms}
\end{eqnarray}
respectively.
Here and in the following, we denote the quantities referring to the
$W^\star N$ CM frame by an asterisk.
The second term on the right-hand side of Eq.~(\ref{eq:ycms}) originates from
the Lorentz boost from the $W^\star a$ CM frame to the $W^\star N$ one.
Here, $y_H^\star$ is taken to be positive in the direction of the
$W^\star$ three-momentum.

We evaluate the cross sections of the partonic subprocesses
(\ref{eq:n})--(\ref{eq:al}) applying the covariant-projector method of
Ref.~\cite{pet}.
In the case of process (\ref{eq:n}), our results can be written in the form
\begin{eqnarray}
\frac{d^3\sigma}{dy\,dQ^2\,d\hat t}(\nu_l+a\to l^-+c\overline{c}[n]+b)
&=&F_a[n]\left[\frac{1+(1-y)^2}{y}T_a[n]-\frac{4(1-y)}{y}L_a[n]\right.
\nonumber\\
&&{}+
\left.\vphantom{\frac{1+(1-y)^2}{y}}
(2-y)A_a[n]\right],
\label{eq:res}
\end{eqnarray}
where $F_a[n]$, $T_a[n]$, $L_a[n]$, and $A_a[n]$ are functions of $\hat s$,
$\hat t$, $\hat u$, and $Q^2$, which are listed in the Appendix.
We combine the results proportional to the CO MEs
$\left\langle{\cal O}^\psi\left[{}^3\!P_J^{(8)}\right]\right\rangle$ and
$\left\langle{\cal O}^{\chi_{cJ}}\left[{}^3\!S_1^{(8)}\right]\right\rangle$
exploiting the multiplicity relations of Eq.~({\ref{eq:mul}).
By charge-conjugation invariance, the results for process (\ref{eq:an})
coincide with Eq.~(\ref{eq:res}).
The results for processes (\ref{eq:l}) and (\ref{eq:al}) may be obtained from
Eq.~(\ref{eq:res}) by scaling $F_a[n]$ with the factor 1/2 to account for the
$l^\pm$ spin average and by flipping the sign of $A_a[n]$ as required by
crossing symmetry.

Invoking the factorization theorems of the QCD parton model and of NRQCD, the
differential cross section of process (\ref{eq:N}) can be written as
\begin{eqnarray}
\lefteqn{\frac{d^2\sigma}{dy\,dQ^2}(\nu_l+N\to l^-+H+j+X)
=\int_{[Q^2+(M+m)^2]/(yS)}^1dx\int_{\hat t_-}^{\hat t_+}d\hat t}
\nonumber\\
&&{}\times
\sum_{a,b,n}f_{a/N}(x,\mu_f)\left\langle{\cal O}^H[n]\right\rangle
\frac{d^3\sigma}{dy\,dQ^2\,d\hat t}(\nu_l+a\to l^-+c\overline{c}[n]+b),
\label{eq:dif}
\end{eqnarray}
where $f_{a/N}(x,\mu_f)$ is the parton density function (PDF) of $a$ in $N$,
$\mu_f$ is the factorization scale,
$\left(d^3\sigma/dy\,dQ^2\,d\hat t\right)(\nu_l+a\to l^-+c\overline{c}[n]+b)$
is given by Eq.~(\ref{eq:res}), and
\begin{equation}
\hat t_\pm=\frac{1}{2\hat s}\left[m^2(\hat s-Q^2)-(\hat s+Q^2)
\left(\hat s-M^2\mp\sqrt{\lambda(\hat s,M^2,m^2)}\right)\right],
\end{equation}
with $\lambda(x,y,z)=x^2+y^2+z^2-2(xy+yz+zx)$ being K\"all\'en's function.
The cross section formulas for processes (\ref{eq:AN})--(\ref{eq:AL}) have the
same form as Eq.~(\ref{eq:dif}).

The kinematically allowed ranges of $S$, $y$, and $Q^2$ are $S>(M+m)^2$,
$(M+m)^2/S<y<1$, and $0<Q^2<yS-(M+m)^2$, respectively.
We then evaluate the cross section of prompt $J/\psi$ production as
\begin{eqnarray}
\lefteqn{d\sigma(\nu_l+N\to l^-+(J/\psi)_{\rm prompt}+j+X)
=d\sigma(\nu_l+N\to l^-+J/\psi+j+X)}
\nonumber\\
&&{}+\sum_{J=0}^2B(\chi_{cJ}\to J/\psi+\gamma)
d\sigma(\nu_l+N\to l^-+\chi_{cJ}+j+X)
\nonumber\\
&&{}+B(\psi^\prime\to J/\psi+X)d\sigma(\nu_l+N\to l^-+\psi^\prime+j+X).
\end{eqnarray}
The kinematic relations for $\nu N$ and $ep$ DIS given in Eqs.~(15)--(24) of
Ref.~\cite{nun} and in Eqs.~(17)--(24) of Ref.~\cite{ep}, respectively, carry
over to the present analysis.
In particular, they are not affected by a finite value of $m$.

\section{Numerical results}
\label{sec:num}

We are now in a position to present our numerical results for prompt $J/\psi$
production in CC DIS.
To be specific, we consider $\nu$Pb scattering at CHORUS and $e^-p$
scattering at HERA and THERA.
The CHORUS experiment uses a wide-band neutrino beam that mainly consists of
$\nu_\mu$ neutrinos, with an average energy of 27~GeV, and contains
$\overline{\nu}_\mu$, $\nu_e$, and $\overline{\nu}_e$ admixtures of about 6\%,
1\%, and below 1\%, respectively.
We evaluate the spectrum-averaged cross section per nucleon bound in lead
using accurate parameterizations for the energy spectra of the individual
neutrino species \cite{nomad} as described in Eq.~(38) of Ref.~\cite{nun}.
Furthermore, we approximate the effective nucleon PDFs of lead using Eq.~(36)
of Ref.~\cite{nun}, with $A=207.2$ and $Z=82$, and include a nuclear
correction factor for incoherent scattering of $1.31{+0.22\atop-0.20}$
\cite{nun}.
At HERA, the lepton and proton energies in the laboratory frame are 27.5~GeV
and 920~GeV, respectively.
The respective values for THERA are planned to be 250~GeV and 1~TeV
\cite{thera}.
We adopt the same input parameters and PDFs, choose the renormalization and
factorization scales, and estimate the theoretical uncertainties as in
Ref.~\cite{nun}.
For simplicity, we take the CKM matrix to be diagonal.

\begin{table}[ht]
\begin{center}
\caption{Integrated cross sections (in fb) of prompt $J/\psi$ production by CC
DIS in $\nu$Pb collisions at CHORUS and in $e^-p$ collisions at HERA and
THERA.
In each case, the NRQCD results for charmless and charmed $X^\prime$ and the
CSM result (for charmed $X^\prime$) are compared.
For comparison, also the respective LO results for fully inclusive CC DIS are
presented.}
\label{tab:xs}
\medskip
\begin{tabular}{|c|cccc|}
\hline\hline
 & charmless & charmed & CSM & inclusive \\
\hline
CHORUS & $4.9{+3.4\atop-2.3}\times10^{-5}$ &
$1.8{+1.5\atop-0.9}\times10^{-5}$ & $1.7{+1.4\atop-0.8}\times10^{-5}$ &
$235\pm39$ \\
HERA & $1.5{+0.5\atop-0.4}$ & $0.50{+0.16\atop-0.13}$ &
$0.43{+0.14\atop-0.11}$ & $70.2{+1.6\atop-0.1}\times10^3$ \\
THERA & $4.5{+1.5\atop-1.2}$ & $2.5{+0.8\atop-0.6}$ &
$2.0{+0.7\atop-0.5}$ & $195{+8\atop-5}\times10^3$ \\
\hline\hline
\end{tabular}
\end{center}
\end{table}
As a first step, we consider the integrated cross sections for CHORUS, HERA, 
and THERA, distinguishing between charmless and charmed underlying events
$X^\prime$ and, in the latter case, between NRQCD and the CSM.
The results are presented in Table~\ref{tab:xs} together with their 
theoretical uncertainties.
For reference, we also quote the respective LO values for fully inclusive CC
DIS, evaluated from Eq.~(14.8) of Ref.~\cite{pdg}, choosing
$\mu_f=\xi_f\sqrt{Q^2}$, where $\xi_f$ is an unphysical scale parameter of
order unity, and otherwise adopting the theoretical input described above.
We estimate the theoretical uncertainties on these values similarly as in
Ref.~\cite{nun}.
We observe that, in the CHORUS, HERA, and THERA experiments, 27\%, 25\%, and
36\%, respectively, of the prompt $J/\psi$ mesons are produced in association
with an additional charmed hadron.
The bulk of such events, namely, 94\%, 86\%, and 80\%, respectively, can be
accounted for by the CSM.
On the other hand, the CSM cannot explain, at LO, $J/\psi$ mesons associated
with charmless underlying events.
Therefore, the observation of such events with the predicted rates would
provide another piece of evidence for the failure of the CSM.
As for CHORUS, the fraction of CC DIS events that contain a prompt $J/\psi$
meson is predicted to be $2.9\times10^{-7}$ in NRQCD and $7.2\times10^{-8}$ in
the CSM.
Both values are compatible with the trimuon rates reported in 
Ref.~\cite{tri}.
As for HERA, the luminosity integrated over the entire operation period is
expected to be of order 1~fb${}^{-1}$, so that there is not much
phenomenological interest in a detailed study of the differential cross 
sections.

In the following, we focus our attention on CHORUS and THERA.
We choose the laboratory frame and take the $J/\psi$ rapidity $y_{J/\psi}$ to
be positive in the flight direction of the incoming lepton.
The $p_T$ and $y_{J/\psi}$ distributions for CHORUS are displayed in
Figs.~\ref{fig:nunpt} and \ref{fig:nuny}, and those for THERA in
Figs.~\ref{fig:eppt} and \ref{fig:epy}, respectively.
In each figure, the NRQCD (solid line) and CSM (dashed line) predictions are
compared.
As explained above, the CSM contributions only arise from Feynman diagrams of
the type shown in Fig.~\ref{fig:fey} (b).

As for CHORUS, the NRQCD and CSM $p_T$ distributions exhibit maxima close to
$p_T=1$~GeV and rapidly fall off with increasing value of $p_T$ because of
limited phase space.
The fall-off is significantly stronger in the CSM because the final state then
always contains an additional heavy quark.
The NRQCD and CSM $y_{J/\psi}$ distributions have their maxima close to 3 and
3.5, respectively.
For $y_{J/\psi}\agt3.5$~GeV, they are similar in shape, while the CSM result
significantly falls short of the NRQCD one as $y_{J/\psi}$ is decreased.

In the case of THERA, the NRQCD and CSM $p_T$ distributions have their maxima
close to $p_T=5$~GeV and smoothly fall off with increasing value of $p_T$.
The NRQCD and CSM $y_{J/\psi}$ distributions have their maxima close to 0 and
1, respectively.
In the forward direction, they are similar in shape, while the CSM result
significantly falls short of the NRQCD one as $y_{J/\psi}$ is taken out to
negative values.

\begin{table}[ht]
\begin{center}
\caption{Fractions of prompt $J/\psi$ mesons in $\nu$Pb CC DIS at CHORUS that
are produced directly and through the feed-down from $\chi_{cJ}$ and
$\psi^\prime$ mesons.
In each case, the NRQCD results for charmless, charmed, and unobserved
$X^\prime$ and the CSM result (for charmed $X^\prime$) are compared.}
\label{tab:fr}
\medskip
\begin{tabular}{|c|cccc|}
\hline\hline
& charmless & charmed & unobserved & CSM \\
\hline
direct & 43\% & 76\% & 52\% & 77\% \\
via $\chi_{cJ}$ & 34\% & 3\% & 26\% & 2\% \\
via $\psi^\prime$ & 23\% & 21\% & 22\% & 21\% \\
\hline\hline
\end{tabular}
\end{center}
\end{table}
It is interesting to study which fractions of the prompt $J/\psi$ mesons are
produced directly and by the various feed-down channels.
This is done for CHORUS in Table~\ref{tab:fr}.
We observe that, within NRQCD, the values for the direct-production and
$\chi_{cJ}$-decay modes significantly differ depending on whether $X^\prime$
is charmless or charmed, while there is hardly any difference in the case of
feed-down from $\psi^\prime$.
As for charmed $X^\prime$, the NRQCD and CSM values are practically
indistinguishable.
However, if $X^\prime$ is unobserved, then NRQCD and the CSM lead to
significantly different results.

\section{Conclusions}
\label{sec:con}

We studied prompt $J/\psi$ production in CC DIS at LO in the NRQCD
factorization formalism, providing the cross sections of all contributing
partonic processes in analytic form.
We presented theoretical predictions for $\nu N$ fixed-target scattering at
CHORUS and for $e^-p$ colliding-beam experiments at HERA and THERA, including
conservative error estimates.

There are two distinct classes of final states, one excluding and one
including charmed hadrons in addition to the $J/\psi$ mesons.
The former one results from pure CO processes, while the latter one may, to a
large extent, already be explained within the CSM.
Observation of $J/\psi$ mesons without any additional charmed hadrons would,
therefore, provide evidence in favour of the CO mechanism.
If the systems accompanying the $J/\psi$ mesons are not observed, then there
are two indicators of the CO mechanism:
(i) the cross section is typically increased by factor of 4 compared to the
CSM; and
(ii) the fraction of $J/\psi$ mesons from $\chi_{cJ}$ decay is dramatically
increased w.r.t.\ the CSM, while the one of direct $J/\psi$ mesons is
substantially reduced.

\bigskip
\noindent
{\bf Acknowledgements}
\smallskip

We thank Vincent Lemaitre for a useful communication regarding
Ref.~\cite{tri}.
The work of L.Z. was supported by the Studienstiftung des deutschen Volkes
through a PhD scholarship.
This work was supported in part by the Deutsche Forschungsgemeinschaft through
Grant No.\ KN~365/1-1, by the Bundesministerium f\"ur Bildung und Forschung
through Grant No.\ 05~HT1GUA/4, and by Sun Microsystems through Academic
Equipment Grant No.~EDUD-7832-000332-GER.

\def\theequation{\Alph{section}.\arabic{equation}}
\begin{appendix}
\setcounter{equation}{0}
\section{Partonic cross sections}

In this Appendix, we present analytic expressions for the coefficients
$F_a[n]$, $T_a[n]$, $L_a[n]$, and $A_a[n]$ appearing in Eq.~(\ref{eq:res}).
In order to compactify the expressions, it is useful to introduce the Lorentz
invariants $s=2q\cdot p$, $t=-2p\cdot p^\prime$, and $u=-2q\cdot p^\prime$,
which are related to the partonic Mandelstam variables by $s=\hat s+Q^2$,
$t=\hat t-m^2$, and $u=\hat u+Q^2-m^2$, respectively.
Furthermore, $g^\prime=2^{3/4}G_F^{1/2}m_W$ and $m_W$ are the gauge coupling
and mass of the $W$ boson, respectively, $G_F$ is Fermi's constant, and
$V_{ud}$ is the CKM matrix.

$\nu_ld(\bar{u})\to l^-c\overline{c}\left[{}^3\!S_1^{(1)}\right]u(\bar{d})$:
\begin{equation}
F=T=L=A=0.
\end{equation}

$\nu_ld(\bar{u})\to l^-c\overline{c}\left[{}^3\!P_0^{(1)}\right]u(\bar{d})$:
\begin{equation}
F=T=L=A=0.
\end{equation}

$\nu_ld(\bar{u})\to l^-c\overline{c}\left[{}^3\!P_1^{(1)}\right]u(\bar{d})$:
\begin{equation}
F=T=L=A=0.
\end{equation}

$\nu_ld(\bar{u})\to l^-c\overline{c}\left[{}^3\!P_2^{(1)}\right]u(\bar{d})$:
\begin{equation}
F=T=L=A=0.
\end{equation}

$\nu_ld(\bar{u})\to l^-c\overline{c}\left[{}^3\!S_0^{(8)}\right]u(\bar{d})$:
\begin{equation}
F=T=L=A=0.
\end{equation}

$\nu_ld(\bar{u})\to l^-c\overline{c}\left[{}^3\!S_1^{(8)}\right]u(\bar{d})$:
\begin{eqnarray}
F & = & \frac{g^{\prime 4} \alpha_s^2 Q^2 |V_{ud}|^2}{72 \pi M^3 s^4 
        (Q^2+m_W^2)^2 (Q^2 - s)^2 (Q^2 - u)^2},\nonumber\\ 
T & = & -2 Q^6 t^2 (2 s + t) 
        -2 Q^4 s [s^2 u - s t (3 t - 2 u)-2 t^3]
        -Q^2 s^2 [s^2 (t - 2 u)+ 2 s (t^2\nonumber\\ 
  &   & - 4 t u - u^2)
        + t (2 t^2 - 2 t u - u^2)]
        -s^3 u (s^2 + 2 s t + 2 t^2 + 2 t u + u^2),\nonumber\\ 
L & = & 2 Q^2 t (Q^2 - s)^2 (s + t)^2,\nonumber\\
A & = &\pm\{ 2 Q^6 s^2 t 
        - 2 Q^4 s t (2 s^2 + s t - t u)
        + Q^2 s^2 [s^2 (t - 2 u)+ 2 s (t^2 + u^2)- t u (2 t - u)]\nonumber\\
  &   & + s^3 u (s - u) (s + 2 t + u)\},
\end{eqnarray}
where the plus (minus) sign refers to the $\nu_l+d$ ($\nu_l+\overline{u}$)
initial state.

$\nu_ld(\bar{u})\to l^-c\overline{c}\left[{}^3\!P_J^{(8)}\right]u(\bar{d})$:
\begin{equation}
F=T=L=A=0.
\end{equation}

$\nu_ls\to l^-c\overline{c}\left[{}^3\!S_1^{(1)}\right]c$:
\begin{eqnarray}
F & = & \frac{g^{\prime 4} \alpha_s^2 |V_{cs}|^2}{27 \pi M s^4 (Q^2 + m_W^2)^2 
        (2 Q^2 + s - 2 t + u)^2 (2 Q^2 - 2 s + t + u)^4},\nonumber\\
T & = & 32 Q^{10} (2 s^3 + 14 s^2 t + 18 s t^2 - 39 t^3) 
        -16 Q^8 [31 s^4- s^3 (133 t + 50 u)+ s^2 t (328 t + u)\nonumber\\
  &   & - 3 s t^2 (33 t-32 u)+ 81 t^3 (t + u)] 
        + 24 Q^6 [46 s^5-s^4 (201 t + 4 u)+2 s^3 (66 t^2 - 97 t u -
        u^2)\nonumber\\ 
  &   & - s^2 t (104 t^2 + 40 t u + 53 u^2)+ 12 s t^2 (t+u) (3 t-7 u)
        - 15 t^3 (t + u)^2] 
        -4 Q^4 [244 s^6\nonumber\\
  &   & -4 s^5 (173 t - 57 u)- 3 s^4 (61 t^2 + 168 t u - 73 u^2)- 
        s^3 (677 t^3 - 72 t^2 u - 591 t u^2 - 154 u^3)\nonumber\\
  &   & + s^2 t (t+u)(52 t^2 - 283 t u + 259 u^2)
        - 9 s t^2 (t + u)^2 (3 t - 16 u) + 3 t^3 (t + u)^3]+ 
        2 Q^2 s [152 s^6\nonumber\\ 
  &   & - 52 s^5 (5 t - 2 u)-2 s^4 (131 t^2 - 16 t u + 93 u^2)
        - s^3 (469 t^3 - 456 t^2 u - 531 t u^2 + 
        238 u^3)\nonumber\\
  &   & + 4 s^2 (t + u)(25 t^3 + 7 t^2 u + 56 t u^2 -25 u^3)
        - s t (t + u)^2 (10 t^2 - 46 t u + 115 u^2)\nonumber\\
  &   & - 6 t^2 u (t + u)^3],\nonumber\\ 
L & = & -16 Q^{10} (4 s^3 - 47 s^2 t + 36 s t^2 - 78 t^3) + 
        16 Q^8 [30 s^4 - s^3 (73 t - 44 u)+ s^2 t (55 t - 38 u)\nonumber\\
  &   & - 3 s t^2 (33 t - 32 u)+ 81 t^3 (t + u)]- 
        4 Q^6 [260 s^5 + s^4 (119 t + 348 u)+ 
        2 s^3 (144 t^2 - 31 t u\nonumber\\
  &   & - 4 u^2)+3 s^2 t (11 t^2 + 166 t u + 12 u^2)
        + 72 s t^2 (t + u) (3 t - 7 u) - 90 t^3 (t + u)^2]\nonumber\\
  &   & + 4 Q^4 [220 s^6+s^5 (391 t + 150 u) 
        + 18 s^4 (23 t^2 - 10 u^2) + 
        s^3 (295 t^3 + 132 t^2 u + 18 t u^2 - 92 u^3)\nonumber\\
  &   & + s^2 t (t + u)(40 t^2 - 271 t u + 175 u^2) 
        - 9 s t^2 (t + u)^2 (3 t - 16 u)+ 3 t^3 (t + u)^3]
        -Q^2 s [240 s^6\nonumber\\
  &   & + 4 s^5 (197 t - 40 u)+ 
        4 s^4 (277 t^2 - 116 t u - 183 u^2)+ 
        3 s^3 (293 t^3 - 110 t^2 u - 143 t u^2 - 104 u^3)\nonumber\\
  &   & +2 s^2 (t + u)(130 t^3 -81 t^2 u + 15 t u^2 + 10 u^3)- 
        s t (t + u)^2 (32 t^2 - 104 t u + 269 u^2)\nonumber\\
  &   & - 12 t^2 u (t + u)^3]- 
        s^2 [16 s^6 - 4 s^5 (29 t - 18 u)-4 s^4 (75 t^2 - 26 t u - 11 u^2)
        - s^3 (215 t^3\nonumber\\
  &   & - 258 t^2 u - 213 t u^2 + 104 u^3) 
        - s^2 (t + u)(47 t^3 - 224 t^2 u - 67 t u^2 + 132 u^3)\nonumber\\
  &   & - s u (t + u)^2 (22 t^2 - 19 t u + 40 u^2)
        - 3 t u^2 (t + u)^3],\nonumber\\ 
A & = & -32 Q^{10} s (15 s^2 - 43 s t + 33 t^2) + 
        16 Q^8 s [105 s^3-s^2 (86 t - 63 u)+ s t (43 t - 38 u) 
        - 9 t^2 (22 t + 9 u)]\nonumber\\  
  &   & - 24 Q^6 s [90 s^4 - 5 s^3 (t - 6 u) + 
        s^2 (5 t + 3 u)(37 t - 5 u)- 
        s t (155 t^2 - 116 t u - 76 u^2) \nonumber\\
  &   & + 21 t^2 (t + u)(3 t + u)] + 
        4 Q^4 s [300 s^5- 8 s^4 (10 t + 9 u) + 
        3 s^3 (389 t^2 - 116 t u - 141 u^2)\nonumber\\
  &   & - 3 s^2 (40 t^3 - 215 t^2 u + 110 t u^2 + 53 u^3)+ 
        s t (t + u)(253 t^2 - 295 t u - 170 u^2)
        - 3 t^2 (t + u)^2 (8 t\nonumber\\
  &   & + 5 u)] - 
        2 Q^2 s^2 [120 s^5 - 100 s^4 t + 
        2 s^3 (289 t^2 - 368 t u - 9 u^2) + 
        3 s^2 (77 t^3 + 62 t^2 u - 95 t u^2\nonumber\\
  &   & + 76 u^3) - 
        2 s (t + u)(2 t^3 + 175 t^2 u - 16 t u^2 - 63 u^3)
        - t (t + u)^2 (4 t^2 - 28 t u - 23 u^2)].
\end{eqnarray}

$\nu_ls\to l^-c\overline{c}\left[{}^3\!P_0^{(1)}\right]c$:
\begin{eqnarray}
F & = & \frac{4 g^{\prime 4} \alpha_s^2 |V_{cs}|^2}{27 \pi M^3 s^4 
        (Q^2 + m_W^2)^2 (2 Q^2 + s - 2 t + u)^4 (2 Q^2 - 2 s + t + u)^5},
        \nonumber\\ 
T & = & 256 Q^{16} t (94 s^2 + 36 s t - 39 t^2) + 
        128 Q^{14} [237 s^4 + 
        s^3 (193 t + 294 u) + s^2 t (58 t + 637 u)\nonumber\\
  &   & - 3 s t^2 (39 t + 8 u)
        + 18 t^3 (8 t - 3 u)] - 
        192 Q^{12} [330 s^5 + s^4 (668 t - 57 u)+ 
        s^3 (883 t^2\nonumber\\
  &   & + 271 t u - 426 u^2) + 
        s^2 t (437 t^2 + 1039 t u - 220 u^2) + 
        3 s t^2 (137 t^2 - 25 t u + 30 u^2)\nonumber\\
  &   & + 3 t^3 (14 t^2 - 11 u^2)] + 
        32 Q^{10} [729 s^6 + s^5 (1805 t - 3492 u) + 
        6 s^4 (1234 t^2 - 1673 t u\nonumber\\
  &   & - 450 u^2) - 
        6 s^3 (108 t^3 + 958 t^2 u + 1907 t u^2 - 132 u^3) + 
        s^2 t (6314 t^3 - 5349 t^2 u - 7635 t u^2\nonumber\\
  &   & - 1885 u^3) - 
        9 s t^2 (101 t^3 + 146 t^2 u + 54 t u^2 + 30 u^3) - 
        12 t^3 (t + u)(38 t^2 + 70 t u - 13 u^2)]\nonumber\\
  &   & + 16 Q^8 [384 s^7 + s^6 (5636 t + 813 u) - 
        s^5 (5617 t^2 - 22685 t u + 6822 u^2)\nonumber\\
  &   & + 6 s^4 (978 t^3 + 1160 t^2 u - 1498 t u^2 - 
        1763 u^3) - 2 s^3 (410 t^4 - 13320 t^3 u + 1257 t^2 u^2\nonumber\\
  &   & + 12083 t u^3 + 1866 u^4) + s^2 t (1568 t^4 + 4703 t^3 u 
        + 1866 t^2 u^2 - 5446 t u^3 - 4690 u^4)\nonumber\\
  &   & + 3 s t^2 (t + u)(339 t^3 + 86 t^2 u - 1166 t u^2 - 40 u^3)
        - 9 t^3 (t + u)^2 (59 t^2 + 50 t u\nonumber\\
  &   & + 9 u^2)] 
        + 24 Q^6 [840 s^8 - 2 s^7 (1954 t - 1545 u) - 
        s^6 (1024 t^2 + 3229 t u - 3243 u^2)\nonumber\\
  &   & - s^5 (2423 t^3 - 536 t^2 u - 7177 t u^2 + 2274 u^3)- 
        s^4 (2897 t^4 + 4662 t^3 u - 7860 t^2 u^2\nonumber\\
  &   & - 1816 t u^3 + 5793 u^4) - 
        s^3 (715 t^5 - 2492 t^4 u - 9804 t^3 u^2 - 3932 t^2 u^3 
        + 4957 t u^4\nonumber\\
  &   & + 2526 u^5) - s^2 t (t + u)(1832 t^4 + 2228 t^3 u - 2727 t^2 u^2 
        - 391 t u^3 + 1247 u^4)\nonumber\\
  &   & + 3 s t^2 (t + u)^2 (286 t^3 + 92 t^2 u - 365 t u^2 - 36 u^3) - 
        6 t^3 (t + u)^3 (64 t^2 - 39 t u + 11 u^2)]\nonumber\\
  &   & - 4 Q^4 [4320 s^9 - 
        4 s^8 (664 t - 1305 u) - 24 s^7 (454 t^2 - 107 t u + 441 u^2) + 
        3 s^6 (1112 t^3\nonumber\\
  &   & + 6299 t^2 u + 5326 t u^2 - 5709 u^3)
        - s^5 (14699 t^4 + 249 t^3 u - 17283 t^2 u^2 - 6325 t u^3\nonumber\\
  &   & - 2196 u^4) - 3 s^4 (2312 t^5 + 1739 t^4 u + 3900 t^3 u^2 + 
        11648 t^2 u^3 + 2576 t u^4 - 4419 u^5)\nonumber\\
  &   & + 3 s^3 (t + u)(1785 t^5 + 6260 t^4 u + 5462 t^3 u^2 - 4586 t^2 u^3 
        - 2543 t u^4 + 1806 u^5)\nonumber\\
  &   & - s^2 t (t + u)^2 (4141 t^4 + 1480 t^3 u + 1815 t^2 u^2 - 1229 t u^3 - 
        1088 u^4)- 
        9 s t^2 (t + u)^3 (192 t^3\nonumber\\
  &   & - 182 t^2 u + 51 t u^2 - 46 u^3)
        + 3 t^3 (t + u)^4 (16 t - 5 u)^2] + 
        2 Q^2 s [336 s^9 + 40 s^8 (184 t - 39 u)\nonumber\\
  &   & - 4 s^7 (1646 t^2 - 2065 t u + 801 u^2) - 
        6 s^6 (1326 t^3 + 2625 t^2 u + 832 t u^2 - 571 u^3)+ 
        s^5 (4835 t^4\nonumber\\
  &   & + 18963 t^3 u - 6609 t^2 u^2 - 10807 t u^3 + 7482 u^4)- 
        2 s^4 (998 t^5 - 7945 t^4 u - 13506 t^3 u^2\nonumber\\
  &   & - 814 t^2 u^3 + 
        3488 t u^4 - 153 u^5) - 6 s^3 (t + u)(473 t^5 + 1925 t^4 u + 1679 t^3
        u^2 - 962 t^2 u^3 - 858 t u^4\nonumber\\
  &   & + 637 u^5)
        + 2 s^2 (t + u)^2 (128 t^5 + 983 t^4 u - 4764 t^3 u^2 - 
        334 t^2 u^3 + 3056 t u^4\nonumber\\
  &   & - 690 u^5)- 
        s t (t + u)^3 (272 t^4 - 5728 t^3 u - 78 t^2 u^2 + 1718 t u^3 + 
        35 u^4)\nonumber\\
  &   & - 6 t^2 u (t + u)^4 (16 t - 5 u)^2],\nonumber\\
L & = & -128 Q^{16} (12 s^3 - 59 s^2 t + 72 s t^2 - 78 t^3) + 
        64 Q^{14} [226 s^4 - 2 s^3 (17 t - 126 u) 
        + s^2 t (431 t - 55 u)\nonumber\\
  &   & + 6 s t^2 (39 t + 8 u) - 36 t^3 (8 t - 3 u)] - 
        32 Q^{12} [1040 s^5 + 10 s^4 (137 t + 59 u)+ 
        12 s^3 (89 t^2\nonumber\\
  &   & + 21 t u - 69 u^2)+ 
        3 s^2 t (1009 t^2 - 652 t u + 245 u^2)- 
        18 s t^2 (137 t^2 - 25 t u + 30 u^2)\nonumber\\
  &   & - 18 t^3 (14 t^2 - 11 u^2)] + 
        16 Q^{10} [1122 s^6 + 2 s^5 (2615 t - 1518 u) - 
        6 s^4 (317 t^2 - 225 t u + 826 u^2)\nonumber\\
  &   & + 12 s^3 (789 t^3 - 1381 t^2 u - 11 t u^2 - 29 u^3) - 
        s^2 t (7547 t^3 - 7545 t^2 u - 2559 t u^2 + 419 u^3)\nonumber\\
  &   & + 18 s t^2 (101 t^3 + 146 t^2 u + 54 t u^2 + 30 u^3) + 
        24 t^3 (t + u)(38 t^2 + 70 t u - 13 u^2)]\nonumber\\
  &   & + 8 Q^8 [3176 s^7 - s^6 (10037 t - 12174 u) + 
        2 s^5 (8093 t^2 - 2803 t u + 2982 u^2) - 
        2 s^4 (10451 t^3\nonumber\\
  &   & - 19740 t^2 u + 336 t u^2 + 3146 u^3) + 
        2 s^3 (10685 t^4 - 5025 t^3 u - 1929 t^2 u^2 + 1145 t u^3\nonumber\\
  &   & - 1602 u^4)- 
        s^2 t (10778 t^4 + 4106 t^3 u - 8970 t^2 u^2 - 4120 t u^3 - 2659 u^4)- 
        6 s t^2 (t + u)(339 t^3\nonumber\\
  &   & + 86 t^2 u - 1166 t u^2 - 40 u^3)
        + 18 t^3 (t + u)^2 (59 t^2 + 50 t u + 9 u^2)]\nonumber\\
  &   & - 4 Q^6 [8944 s^8 - 8 s^7 (1612 t - 1673 u) + 
        3 s^6 (9355 t^2 - 6847 t u - 3274 u^2) - 
        2 s^5 (12211 t^3\nonumber\\
  &   & - 23634 t^2 u + 3837 t u^2 + 10282 u^3) + 
        2 s^4 (13886 t^4 - 9175 t^3 u - 9240 t^2 u^2 + 212 t u^3\nonumber\\
  &   & - 2017 u^4) + 
        6 s^3 (943 t^5 + 5888 t^4 u + 1572 t^3 u^2 - 4856 t^2 u^3 - 1069 t u^4 +
        378 u^5)\nonumber\\
  &   & - 3 s^2 t (t + u)(3608 t^4 - 268 t^3 u - 5828 t^2 u^2 + 1462 t u^3 +
        1785 u^4)\nonumber\\
  &   & + 18 s t^2 (t + u)^2 
        (286 t^3 + 92 t^2 u - 365 t u^2 - 36 u^3) - 36 t^3 (t + u)^3 
        (64 t^2 - 39 t u + 11 u^2)]\nonumber\\
  &   & + 2 Q^4 [6912 s^9 - 8 s^8 (1121 t - 408 u) + 
        6 s^7 (2257 t^2 - 4629 t u - 3094 u^2)- 
        3 s^6 (655 t^3\nonumber\\
  &   & - 12272 t^2 u + 779 t u^2 + 3226 u^3)
        + 2 s^5 (3611 t^4 - 12591 t^3 u - 4287 t^2 u^2 + 
        12773 t u^3\nonumber\\
  &   & + 8490 u^4) + 6 s^4 (5158 t^5 + 5319 t^4 u - 7411 t^3 u^2 - 
        7604 t^2 u^3 + 2031 t u^4 + 2315 u^5)\nonumber\\
  &   & - 6 s^3 (t + u)(2059 t^5 + 892 t^4 u + 804 t^3 u^2 - 186 t^2 u^3 
        - 850 t u^4 - 358 u^5)\nonumber\\
  &   & + s^2 t (t + u)^2 (6110 t^4 + 7808 t^3 u - 16386 t^2 u^2 + 1526 t u^3 +
        4139 u^4)- 18 s t^2 (t + u)^3 (192 t^3\nonumber\\
  &   & - 182 t^2 u + 51 t u^2 - 46 u^3) + 6 t^3 (t + u)^4 (16 t - 5 u)^2] - 
        Q^2 s [992 s^9 - 32 s^8 (55 t + 39 u)\nonumber\\
  &   & - 4 s^7 (629 t^2 + 2255 t u + 762 u^2) + 
        4 s^6 (1480 t^3 + 141 t^2 u - 1146 t u^2 + 1873 u^3)
        + s^5 (173 t^4\nonumber\\
  &   & - 9 t^3 u + 1887 t^2 u^2 + 6293 t u^3 + 
        10272 u^4)+ 
        4 s^4 (2459 t^5 - 3733 t^4 u - 5061 t^3 u^2 + 2339 t^2 u^3\nonumber\\
  &   & - 490 t u^4 - 
        1266 u^5) + 2 s^3 (t + u)(4822 t^5 - 928 t^4 u - 8213 t^3 u^2 + 12406
        t^2 u^3 + 1091 t u^4\nonumber\\
  &   & - 5084 u^5)
        - 4 s^2 (t + u)^2 (716 t^5 - 295 t^4 u + 2937 t^3 u^2 - 
        3040 t^2 u^3 - 616 t u^4\nonumber\\
  &   & + 783 u^5)+ 
        s t (t + u)^3 (3568 t^4 + 592 t^3 u - 876 t^2 u^2 + 1384 t u^3 - 
        1391 u^4)\nonumber\\
  &   & - 12 t^2 u (t + u)^4 (16 t - 5 u)^2] - 
        s^2 [64 s^9 + 8 s^8 (41 t + 38 u) + 
        12 s^7 (49 t^2 + 67 t u + 10 u^2)\nonumber\\
  &   & + 2 s^6 (103 t^3 + 174 t^2 u - 201 t u^2 - 248 u^3) - 
        s^5 (169 t^4 + 283 t^3 u + 1995 t^2 u^2 + 1441 t u^3\nonumber\\
  &   & - 152 u^4)
        - 3 s^4 (137 t^5 - 240 t^4 u + 494 t^3 u^2 + 896 t^2 u^3 - 
        351 t u^4 - 304 u^5)- 
        2 s^3 (t + u)(820 t^5\nonumber\\
  &   & - 922 t^4 u + 49 t^3 u^2 + 1420 t^2 u^3 
        - 1441 t u^4 +172 u^5) - 2 s^2 (t + u)^2 (640 t^5\nonumber\\
  &   & - 1288 t^4 u - 
        239 t^3 u^2 + 1396 t^2 u^3 - 1540 t u^4 + 616 u^5)+ 
        3 s u (t + u)^3 (464 t^4 - 216 t^3 u\nonumber\\
  &   & - 180 t^2 u^2 + 377 t u^3 - 
        168 u^4) - 3 t u^2 (t + u)^4 (16 t - 5 u)^2],\nonumber\\
A & = & 256 Q^{16} s (35 s^2 + 107 s t + 141 t^2) - 
        128 Q^{14} s [115 s^3 - s^2 (t + 260 u) + s t (862 t - 839 u)\nonumber\\
  &   & + 81 t^2 (5 t - 2 u)] 
        + 192 Q^{12} s [143 s^4 - 115 s^3 (3 t - u) + 
        s^2 (192 t^2 - 901 t u + 359 u^2)\nonumber\\
  &   & - s t (178 t^2 + 1529 t u - 269 u^2)- 
        3 t^2 (13 t^2 + 101 t u + 61 u^2)] - 
        32 Q^{10} s [2141 s^5 + s^4 (329 t\nonumber\\
  &   & + 3516 u) - 
        6 s^3 (255 t^2 - 537 t u - 205 u^2) + 
        2 s^2 (826 t^3 + 765 t^2 u + 6279 t u^2 - 437 u^3)\nonumber\\
  &   & - s t (2710 t^3 - 33 t^2 u - 7305 t u^2 - 2645 u^3) + 
        3 t^2 (485 t^3 + 384 t^2 u + 258 t u^2 + 332 u^3)]\nonumber\\
  &   & + 16 Q^8 s [3304 s^6 + s^5 (7388 t - 515 u) + 
        s^4 (559 t^2 + 11005 t u - 10758 u^2) + 
        2 s^3 (3551 t^3\nonumber\\
  &   & + 1506 t^2 u - 2277 t u^2 - 5281 u^3) + 
        s^2 (4661 t^4 + 14932 t^3 u - 108 t^2 u^2 - 19618 t u^3\nonumber\\
  &   & - 3299 u^4) + 
        s t (1783 t^4 + 2947 t^3 u + 3624 t^2 u^2 - 3722 t u^3 - 5615 u^4) - 
        27 t^2 (t + u)(4 t^3\nonumber\\
  &   & + 87 t^2 u + 27 t u^2 + 7 u^3)] + 
        24 Q^6 s [440 s^7 - 2 s^6 (1866 t - 2009 u) - 
        s^5 (2304 t^2 + 157 t u\nonumber\\
  &   & - 4715 u^2) - 
        s^4 (3041 t^3 + 290 t^2 u - 4019 t u^2 + 2728 u^3) - 
        s^3 (3933 t^4 - 430 t^3 u - 6696 t^2 u^2\nonumber\\
  &   & + 1020 t u^3 + 6377 u^4) - 
        s^2 (351 t^5 - 3614 t^4 u - 9248 t^3 u^2 - 4764 t^2 u^3 + 3517 t u^4 + 
        2512 u^5)\nonumber\\
  &   & - s t (t + u)(2140 t^4 + 1426 t^3 u + 51 t^2 u^2 - 989 t u^3 
        + 1081 u^4) + 3 t^2 (t + u)^2(4 t - 5 u)(20 t^2\nonumber\\
  &   & + 37 t u - 10 u^2)] - 
        4 Q^4 s [4336 s^8 - 4 s^7 (332 t - 1761 u) - 
        24 s^6 (488 t^2 - 207 t u\nonumber\\
  &   & + 383 u^2) + 
        s^5 (2936 t^3 + 14697 t^2 u + 7338 t u^2 - 19759 u^3) 
        - s^4 (15136 t^4 - 9795 t^3 u\nonumber\\
  &   & - 19941 t^2 u^2 - 2225 t u^3 - 
        303 u^4) - 3 s^3 (2535 t^5 + 2685 t^4 u + 4606 t^3 u^2 + 
        8496 t^2 u^3\nonumber\\
  &   & - 57 t u^4 - 4637 u^5) + 
        s^2 (t + u)(2488 t^5 + 21065 t^4 u + 4390 t^3 u^2 - 13118 t^2 u^3 -
        6790 t u^4\nonumber\\
  &   & + 5749 u^5) - 
        s t (t + u)^2 (4952 t^4 - 1612 t^3 u - 933 t^2 u^2 - 901 t u^3 + 
        29 u^4)\nonumber\\
  &   & + 3 t^2 (t + u)^3 (352 t^3 - 528 t^2 u + 372 t u^2 - 71 u^3)]
        + 2 Q^2 s^2 [368 s^8 + 8 s^7 (956 t - 179 u)\nonumber\\
  &   & - 4 s^6 (1510 t^2 - 2033 t u + 921 u^2) - 
        2 s^5 (3286 t^3 + 7503 t^2 u + 3348 t u^2 - 1085 u^3)\nonumber\\
  &   & + s^4 (3713 t^4 + 19387 t^3 u - 2925 t^2 u^2 - 9199 t u^3 + 
        7240 u^4)- 
        4 s^3 (1274 t^5 - 1579 t^4 u\nonumber\\
  &   & - 4845 t^3 u^2 - 1150 t^2 u^3 + 
        737 t u^4 - 267 u^5) - 2 s^2 (t + u)(932 t^5 + 2224 t^4 u + 7253 t^3
        u^2\nonumber\\
  &   & - 1354 t^2 u^3 - 2087 t u^4 + 1826 u^5)
        + 2 s (t + u)^2 (800 t^5 + 3504 t^4 u - 3020 t^3 u^2 - 
        322 t^2 u^3\nonumber\\
  &   & + 2448 t u^4 - 767 u^5) - 
        t (t + u)^3 (784 t^4 + 544 t^3 u - 1992 t^2 u^2 + 1840 t u^3 - 
        377 u^4)].
\end{eqnarray}

$\nu_ls\to l^-c\overline{c}\left[{}^3\!P_1^{(1)}\right]c$:
\begin{eqnarray}
F & = & \frac{-8 g^{\prime 4} \alpha_s^2 |V_{cs}|^2}{27 \pi M^3 s^4 
        (Q^2 + m_W^2)^2 (2 Q^2 + s - 2 t + u)^4 (2 Q^2 - 2 s + t + u)^5},
        \nonumber\\
T & = & 256 Q^{16} (2 s^3 + 18 s^2 t - 57 s t^2 + 78 t^3) - 
        128 Q^{14} [4 s^4 + s^3 (352 t + 13 u) 
        - 3 s^2 t (53 t - 60 u)\nonumber\\
  &   & + 3 s t^2 (108 t - 19 u) - 9 t^3 (16 t + 29 u)] - 
        192 Q^{12} [28 s^5 - s^4 (451 t - 53 u) + s^3 (67 t^2 
        + 152 t u\nonumber\\
  &   & + 97 u^2)- 
        3 s^2 t (261 t^2 + 104 t u - 35 u^2) + 
        6 s t^2 (17 t^2 + 19 t u - 40 u^2) - 
        3 t^3 (28 t^2 + 55 t u \nonumber\\
  &   & + 33 u^2)] + 
        32 Q^{10} [466 s^6 - 3 s^5 (467 t - 247 u) + 
        3 s^4 (209 t^2 + 1163 t u + 46 u^2) - 
        s^3 (2735 t^3\nonumber\\
  &   & - 4422 t^2 u - 3189 t u^2 + 488 u^3)- 
        27 s^2 t (10 t^3 - 153 t^2 u - 100 t u^2 - 35 u^3)- 
        3 s t^2 (176 t^3\nonumber\\
  &   & + 81 t^2 u - 381 t u^2 - 412 u^3)+ 
        3 t^3 (t + u)(106 t^2 + 161 t u + 46 u^2)] - 
        16 Q^8 [929 s^7\nonumber\\
  &   & + s^6 (1181 t + 539 u) + 
        3 s^5 (735 t^2 + 1277 t u - 680 u^2) + 
        s^4 (2573 t^3 + 6228 t^2 u - 4521 t u^2\nonumber\\
  &   & - 2440 u^3) + s^3 (3920 t^4 + 11969 t^3 u - 6441 t^2 u^2 - 8791 t u^3 - 
        628 u^4)- 3 s^2 t (148 t^4 - 498 t^3 u\nonumber\\
  &   & + 201 t^2 u^2 + 1405 t u^3 + 
        900 u^4)+ 3 s t^2 (t + u)(225 t^3 + 55 t^2 u - 454 t u^2 - 293 u^3)
        \nonumber\\
  &   & - 9 t^3 (t + u)^2 (22 t^2 + 21 t u + 8 u^2)] 
        + 24 Q^6 [242 s^8 + 2 s^7 (485 t - 109 u) + 
        s^6 (1373 t^2 - 467 t u\nonumber\\
  &   & - 772 u^2) 
        + s^5 (1709 t^3 - 1833 t^2 u - 2817 t u^2 + 629 u^3) + 
        s^4 (2965 t^4 + 1147 t^3 u - 5271 t^2 u^2\nonumber\\
  &   & + 487 t u^3 + 1654 u^4)
        + s^3 (107 t^5 - 1925 t^4 u - 5218 t^3 u^2 - 778 t^2 u^3 + 
        2653 t u^4 + 713 u^5)\nonumber\\
  &   & + 3 s^2 t (t + u)(303 t^4 + 134 t^3 u - 250 t^2 u^2 + 118 t u^3 
        + 262 u^4) - 3 s t^2 (t + u)^3 
        (22 t^2 - 22 t u\nonumber\\
  &   & - 41 u^2) + 3 t^3 (t + u)^3 (22 t^2 - 20 t u + 
        9 u^2)] - 4 Q^4 [76 s^9 + 12 s^8 (113 t - 49 u) + 
        3 s^7 (879 t^2\nonumber\\
  &   & - 1226 t u - u^2)+ 
        s^6 (2963 t^3 - 8481 t^2 u - 1443 t u^2 + 1745 u^3) + 
        3 s^5 (695 t^4 - 4291 t^3 u\nonumber\\
  &   & - 1113 t^2 u^2 + 2915 t u^3 - 346 u^4)
        + 3 s^4 (2320 t^5 + 493 t^4 u - 250 t^3 u^2 + 4396 t^2 u^3 + 
        1150 t u^4\nonumber\\
  &   & - 1309 u^5) + s^3 (t + u)(847 t^5 - 5620 t^4 u - 566 t^3 u^2 
        + 5962 t^2 u^3 - 664 t u^4 - 1805 u^5)\nonumber\\
  &   & + 9 s^2 t (t + u)^2 (119 t^4 - 88 t^3 u - 39 t^2 u^2 + 29 t u^3 - 
        85 u^4) + 3 s t^2 (t + u)^3 (4 t^3 - 138 t^2 u\nonumber\\
  &   & + 111 t u^2 - 62 u^3) - 33 t^3 (t + u)^4 (2 t - u)^2] - 
        2 Q^2 s [88 s^9 + 8 s^8 (17 t + 23 u) + 
        6 s^7 (7 t^2\nonumber\\
  &   & + 136 t u - 7 u^2) - 4 s^6 (265 t^3 - 588 t^2 u + 
        237 t u^2 - 14 u^3) - s^5 (499 t^4 - 2141 t^3 u 
        + 5307 t^2 u^2\nonumber\\
  &   & + 1393 t u^3 - 434 u^4)- 6 s^4 (180 t^5 - 937 t^4 u + 94 t^3 u^2 
        + 768 t^2 u^3 - 454 t u^4 + 61 u^5)\nonumber\\
  &   & - 2 s^3 (t + u)(1411 t^5 - 169 t^4 u + 769 t^3 u^2 - 296 t^2 u^3 
        - 1819 t u^4 + 556 u^5)\nonumber\\
  &   & + 2 s^2 (t + u)^2 
        (452 t^5 + 889 t^4 u - 658 t^3 u^2 - 178 t^2 u^3 + 502 t u^4 - 
        253 u^5)- 3 s t (t + u)^3 (32 t^4\nonumber\\
  &   & - 120 t^3 u + 150 t^2 u^2 + 2 t u^3 + 15 u^4) 
        - 66 t^2 u (t + u)^4 (2 t - u)^2],\nonumber\\
L & = & 128 Q^{16} (21 s^3 - 52 s^2 t + 114 s t^2 - 156 t^3) - 
        64 Q^{14} [196 s^4 - 2 s^3 (142 t - 31 u) + 5 s^2 t (179 t\nonumber\\
  &   & - 7 u) 
        - 6 s t^2 (108 t - 19 u) + 18 t^3 (16 t + 29 u)] + 
        32 Q^{12} [740 s^5 - 2 s^4 (314 t - 75 u) + 
        s^3 (2217 t^2\nonumber\\
  &   & - 956 t u - 401 u^2) + 
        6 s^2 t (76 t^2 + 68 t u + 53 u^2) + 
        36 s t^2 (17 t^2 + 19 t u - 40 u^2) - 18 t^3 (28 t^2\nonumber\\
  &   & + 55 t u + 33 u^2)] 
        - 16 Q^{10} [1428 s^6 - 4 s^5 (259 t + 123 u) + 
        6 s^4 (58 t^2 - 689 t u - 293 u^2)\nonumber\\
  &   & + 6 s^3 (586 t^3 - 609 t^2 u + 213 t u^2 + 42 u^3)+ 
        s^2 t (317 t^3 - 2109 t^2 u - 4809 t u^2 + 281 u^3)\nonumber\\
  &   & - 6 s t^2 (176 t^3 + 81 t^2 u - 381 t u^2 - 412 u^3) + 
        6 t^3 (t + u)(106 t^2 + 161 t u + 46 u^2)]\nonumber\\
  &   & + 8 Q^8 [1429 s^7 - 2 s^6 (526 t + 827 u) 
        - 2 s^5 (3841 t^2 + 2623 t u + 855 u^2) - 
        2 s^4 (151 t^3\nonumber\\
  &   & + 6021 t^2 u - 3543 t u^2 - 1217 u^3)- 
        s^3 (1718 t^4 + 7180 t^3 u + 1764 t^2 u^2 - 3364 t u^3 
        - 791 u^4)\nonumber\\
  &   & + 2 s^2 t (116 t^4 + 107 t^3 u + 1245 t^2 u^2 + 1166 t u^3 - 925 u^4) + 
        6 s t^2 (t + u)(225 t^3 + 55 t^2 u - 454 t u^2\nonumber\\
  &   & - 293 u^3)
        - 18 t^3 (t + u)^2 (22 t^2 + 21 t u + 8 u^2)] - 
        4 Q^6 [584 s^8 + 2 s^7 (847 t - 489 u)\nonumber\\
  &   & - s^6 (13833 t^2 - 2389 t u - 1102 u^2) - 
        8 s^5 (1412 t^3 + 1143 t^2 u - 1827 t u^2 - 430 u^3)\nonumber\\
  &   & - 2 s^4 (3692 t^4 - 615 t^3 u - 9552 t^2 u^2 - 2788 t u^3 + 117 u^4) - 
        2 s^3 (1581 t^5 + 4291 t^4 u - 1571 t^3 u^2\nonumber\\
  &   & - 3201 t^2 u^3 + 1801 t u^4 + 
        505 u^5) + 3 s^2 t (t + u)(10 t^4 - 1160 t^3 u - 694 t^2 u^2 
        + 662 t u^3 + 789 u^4)\nonumber\\ 
  &   & - 18 s t^2 (t + u)^3 
        (22 t^2 - 22 t u - 41 u^2) + 18 t^3 (t + u)^3 
        (22 t^2 - 20 t u + 9 u^2)]\nonumber\\
  &   & - 2 Q^4 [72 s^9 - 44 s^8 (127 t + 3 u) + 
        3 s^7 (2743 t^2 - 3472 t u - 247 u^2)+ 
        12 s^6 (1199 t^3 + 261 t^2 u\nonumber\\
  &   & - 471 t u^2 + 139 u^3)
        + 2 s^5 (5731 t^4 - 9729 t^3 u - 9186 t^2 u^2 + 5461 t u^3 + 
        2391 u^4) + 6 s^4 (1729 t^5\nonumber\\
  &   & - 92 t^4 u - 4442 t^3 u^2 - 530 t^2 u^3 + 
        2035 t u^4 + 448 u^5) + 3 s^3 (t + u)(978 t^5 + 1200 t^4 u - 1464 t^3
        u^2\nonumber\\
  &   & + 604 t^2 u^3 + 581 t u^4 + 37 u^5)- 
        2 s^2 t (t + u)^2 (205 t^4 + 352 t^3 u + 1428 t^2 u^2 
        - 548 t u^3\nonumber\\
  &   & - 695 u^4) - 6 s t^2 (t + u)^3 (4 t^3 - 138 t^2 u + 111 t u^2 - 62 u^3)
        + 66 t^3 (t + u)^4 (2 t - u)^2] + Q^2 s [32 s^9\nonumber\\
  &   & - 56 s^8 (83 t - 5 u) 
        - 4 s^7 (214 t^2 + 1993 t u - 333 u^2) + 
        8 s^6 (785 t^3 + 363 t^2 u + 195 t u^2 + 209 u^3)\nonumber\\
  &   & + s^5 (8713 t^4 - 12949 t^3 u + 5931 t^2 u^2 + 8161 t u^3 - 1288 u^4)+ 
        4 s^4 (2149 t^5 - 3863 t^4 u - 4827 t^3 u^2\nonumber\\
  &   & + 3439 t^2 u^3 + 364 t u^4 - 1026 u^5)+ 
        2 s^3 (t + u)(1810 t^5 - 2224 t^4 u - 1073 t^3 u^2 + 6574 t^2
        u^3\nonumber\\ 
  &   & - 463 t u^4 - 1430 u^5)
        + 4 s^2 (t + u)^2 (224 t^5 - 207 t^4 u - 539 t^3 u^2 + 
        740 t^2 u^3 - 282 t u^4 - 158 u^5)\nonumber\\
  &   & + s t (t + u)^3 (344 t^4 - 592 t^3 u - 60 t^2 u^2 + 584 t u^3 - 427 u^4)
        - 132 t^2 u (t + u)^4 (2 t - u)^2]\nonumber\\
  &   & + s^2 [40 s^9 + 8 s^8 (83 t + 9 u) + 
        2 s^7 (375 t^2 + 436 t u - 11 u^2) - 16 s^6 (7 t^3 + 69 t^2 u + 51 t u^2
        + u^3)\nonumber\\
  &   & - s^5 (1171 t^4 - 129 t^3 u + 4071 t^2 u^2 + 469 t u^3 - 
        78 u^4) - s^4 (1545 t^5 - 2824 t^4 u - 934 t^3 u^2\nonumber\\
  &   & + 1440 t^2 u^3 - 
        2179 t u^4 + 248 u^5)- 
        2 s^3 (t + u)(400 t^5 - 1450 t^4 u + 79 t^3 u^2 + 646 t^2 u^3 - 1336 t
        u^4\nonumber\\
  &   & + 325 u^5) - 2 s^2 (t + u)^2 (94 t^5 - 518 t^4 u + 
        489 t^3 u^2 + 574 t^2 u^3 - 611 t u^4 + 240 u^5)\nonumber\\
  &   & + s u (t + u)^3 (80 t^4 - 184 t^3 u + 12 t^2 u^2 + 185 t u^3 - 
        118 u^4) - 33 t u^2 (t + u)^4 (2 t - u)^2],\nonumber\\
A & = & -768 Q^{16} s (s^2 + 10 s t + 14 t^2) + 
        384 Q^{14} s [11 s^3 + s^2 (54 t + u) - s t (20 t + 107 u) 
        + 3 t^2 (60 t + 7 u)]\nonumber\\
  &   & - 576 Q^{12} s [17 s^4 + 2 s^3 (44 t + 13 u) 
        - 2 s^2 (41 t^2 + 36 t u - 14 u^2)
        + 2 s t (65 t^2 - 73 t u + 15 u^2)\nonumber\\
  &   & - t^2 (76 t^2 + 202 t u + 69 u^2)] + 
        96 Q^{10} s [112 s^5 + s^4 (943 t + 237 u) - 
        s^3 (187 t^2 - 102 t u + 39 u^2)\nonumber\\
  &   & + s^2 (1213 t^3 + 552 t^2 u + 957 t u^2 - 182 u^3) - 
        s t (782 t^3 + 711 t^2 u - 1560 t u^2 - 481 u^3) + 
        t^2 (236 t^3\nonumber\\
  &   & + 603 t^2 u + 759 t u^2 + 338 u^3)] - 
        48 Q^8 s [83 s^6 + s^5 (1527 t - 47 u) 
        + s^4 (1195 t^2 - 35 t u - 852 u^2)\nonumber\\
  &   & + s^3 (2387 t^3 + 57 t^2 u - 1557 t u^2 - 835 u^3) + 
        s^2 (951 t^4 + 805 t^3 u + 153 t^2 u^2 - 2199 t u^3 
        - 167 u^4)\nonumber\\
  &   & + s t (664 t^4 + 1522 t^3 u + 405 t^2 u^2 - 1865 t u^3 - 1034 u^4) - 
        3 t^2 (t + u)(72 t^3 + 118 t^2 u + 83 t u^2\nonumber\\
  &   & + 64 u^3)] - 
        72 Q^6 s [26 s^7 - 2 s^6 (127 t - 103 u) - 
        s^5 (919 t^2 - 345 t u - 268 u^2) - 
        s^4 (645 t^3 - 943 t^2 u\nonumber\\
  &   & - 741 t u^2 + 271 u^3) - 
        2 s^3 (540 t^4 - 440 t^3 u - 495 t^2 u^2 + 205 t u^3 + 299 u^4) - 
        s^2 (361 t^5 - 279 t^4 u\nonumber\\
  &   & - 1294 t^3 u^2 - 482 t^2 u^3 + 735 t u^4 + 
        239 u^5) - s t (t + u)(62 t^4 - 12 t^3 u - 108 t^2 u^2 + 68 t u^3 
        + 237 u^4)\nonumber\\
  &   & - t^2 (t + u)^2 (46 t^3 + 26 t^2 u + 37 t u^2 + 3 u^3)] 
        + 12 Q^4 s [140 s^8 + 4 s^7 (83 t + 93 u)\nonumber\\
  &   & - s^6 (1487 t^2 - 1290 t u + 255 u^2) - 
        s^5 (1337 t^3 - 1623 t^2 u + 303 t u^2 + 959 u^3)- 
        s^4 (1124 t^4\nonumber\\
  &   & - 7059 t^3 u - 903 t^2 u^2 + 2507 t u^3 - 273 u^4) - 
        s^3 (2821 t^5 + 663 t^4 u + 900 t^3 u^2 + 3844 t^2 u^3\nonumber\\
  &   & + 483 t u^4 - 
        1383 u^5) + s^2 (t + u)(224 t^5 + 2029 t^4 u - 154 t^3 u^2 - 1702 t^2
        u^3 + 247 t u^4 + 638 u^5)\nonumber\\
  &   & - s t (t + u)^2 (484 t^4 - 320 t^3 u - 117 t^2 u^2 + 79 t u^3 - 122 u^4)
        + t^2 (t + u)^3 (124 t^3 - 84 t^2 u\nonumber\\
  &   & + 72 t u^2 - 17 u^3)] 
        - 6 Q^2 s^2 [40 s^8 + 8 s^7 (39 t + u) + 
        2 s^6 (151 t^2 + 208 t u - 111 u^2) - 
        4 s^5 (205 t^3\nonumber\\
  &   & + 12 t^2 u + 81 t u^2 + 34 u^3) - 
        s^4 (663 t^4 - 643 t^3 u + 1881 t^2 u^2 - 129 t u^3 - 220 u^4)
        - 4 s^3 (29 t^5\nonumber\\
  &   & - 757 t^4 u - 57 t^3 u^2 + 
        317 t^2 u^3 - 298 t u^4 + 6 u^5) - 
        2 s^2 (t + u)(376 t^5 - 118 t^4 u + 379 t^3 u^2\nonumber\\
  &   & + 76 t^2 u^3 
        - 448 t u^4 + 187 u^5)+ 
        4 s (t + u)^2 (60 t^5 + 122 t^4 u - 115 t^3 u^2 + 21 t^2 u^3 + 
        71 t u^4 - 46 u^5)\nonumber\\
  &   & - t (t + u)^3 (40 t^4 + 112 t^3 u - 96 t^2 u^2 + 100 t u^3 - 29 u^4)].
\end{eqnarray}

$\nu_ls\to l^-c\overline{c}\left[{}^3\!P_2^{(1)}\right]c$:
\begin{eqnarray}
F & = & \frac{-8 g^{\prime 4} \alpha_s^2 |V_{cs}|^2}{135 \pi M^3 s^4 
        (Q^2 + m_W^2)^2 (2 Q^2 + s - 2 t + u)^4 (2 Q^2 - 2 s + t + u)^5},
        \nonumber\\ 
T & = & 256 Q^{16} (30 s^3 + 2 s^2 t + 117 s t^2 - 384 t^3) - 
        128 Q^{14} [306 s^4 - s^3 (608 t + 87 u) + s^2 t 
        (2167 t - 62 u)\nonumber\\
  &   & - 3 s t^2 (336 t - 337 u) - 9 t^3 (188 t - 57 u)] + 
        192 Q^{12} [432 s^5 - s^4 (859 t - 39 u) + 
        s^3 (421 t^2\nonumber\\
  &   & - 1574 t u - 99 u^2) + 
        s^2 t (3755 t^2 + 1186 t u - 223 u^2) - 12 s t^2 (151 t^2 - 200 t u + 
        60 u^2) + 3 t^3 (272 t^2\nonumber\\
  &   & + 729 t u + 217 u^2)] - 
        32 Q^{10} [2784 s^6 - s^5 (1573 t - 999 u) - 
        3 s^4 (6355 t^2 + 2101 t u - 102 u^2)\nonumber\\
  &   & + 3 s^3 (7219 t^3 - 8714 t^2 u + 875 t u^2 + 292 u^3)- 
        s^2 t (6874 t^3 + 6951 t^2 u + 26382 t u^2 - 911 u^3)\nonumber\\
  &   & + 9 s t^2 (444 t^3 + 475 t^2 u - 2109 t u^2 - 460 u^3) + 
        3 t^3 (t + u)(346 t^2 - 1507 t u - 1754 u^2)]\nonumber\\
  &   & + 16 Q^8 [2673 s^7 + s^6 (6973 t - 1239 u) - 
        s^5 (44033 t^2 - 13345 t u + 4104 u^2) - 3 s^4 (3193 t^3\nonumber\\
  &   & + 10588 t^2 u - 
        12485 t u^2 + 228 u^3) - s^3 (27296 t^4 + 39099 t^3 u - 
        21243 t^2 u^2 - 26065 t u^3 - 156 u^4)\nonumber\\
  &   & - s^2 t (3176 t^4 + 11324 t^3 u + 15915 t^2 u^2 - 37489 t u^3 - 
        4216 u^4) + 3 s t^2 (t + u)(1551 t^3 - 2225 t^2 u\nonumber\\
  &   & + 3494 t u^2 + 4615
        u^3) - 9 t^3 (t + u)^2 (52 t^2 + 277 t u - 522 u^2)]
        + 24 Q^6 [126 s^8 - 2 s^7 (2477 t\nonumber\\
  &   & - 1167 u)+ 
        s^6 (12919 t^2 - 6887 t u + 3942 u^2) + 
        s^5 (9191 t^3 - 9857 t^2 u - 3607 t u^2 + 2889 u^3)\nonumber\\
  &   & + s^4 (16733 t^4 - 11163 t^3 u - 34293 t^2 u^2 + 9293 t u^3 + 1848 u^4)
        + s^3 (10549 t^5 - 521 t^4 u\nonumber\\
  &   & - 15282 t^3 u^2 - 11090 t^2 u^3 + 
        12535 t u^4 + 693 u^5) - s^2 t (t + u)(1033 t^4 - 2882 t^3 u 
        + 7380 t^2 u^2\nonumber\\
  &   & - 3908 t u^3 - 3512 u^4) + 3 s t^2 (t + u)^2 
        (62 t^3 + 460 t^2 u - 1051 t u^2 + 1413 u^3)\nonumber\\
  &   & + 3 t^3 (t + u)^3 (134 t^2 - 312 t u + 187 u^2)] - 
        4 Q^4 [2868 s^9 - 4 s^8 (3491 t - 1809 u) + 
        3 s^7 (3113 t^2\nonumber\\
  &   & - 5398 t u - 87 u^2) + 3 s^6 (3359 t^3 - 5663 t^2 u + 
        8093 t u^2 - 5429 u^3)+ s^5 (22193 t^4 - 83613 t^3 u\nonumber\\
  &   & - 6195 t^2 u^2 + 
        50309 t u^3 - 19170 u^4)+ 
        3 s^4 (19832 t^5 - 11011 t^4 u - 8238 t^3 u^2 
        + 17540 t^2 u^3\nonumber\\
  &   & + 6170 t u^4 - 3165 u^5) + 3 s^3 (t + u)(1493 t^5 - 2368 t^4 u - 4062
        t^3 u^2 + 14602 t^2 u^3 - 3930 t u^4\nonumber\\
  &   & - 661 u^5)
        - s^2 t (t + u)^2 (5225 t^4 - 2788 t^3 u + 
        10263 t^2 u^2 - 10177 t u^3 + 8483 u^4)\nonumber\\
  &   & + 9 s t^2 (t + u)^3 (196 t^3 
        - 662 t^2 u + 909 t u^2 - 486 u^3)
        - 3 t^3 (t + u)^4 
        (88 t^2 - 136 t u + 55 u^2)]\nonumber\\
  &   & + 2 Q^2 s [1656 s^9 - 8 s^8 (767 t - 381 u) - 2 s^7 (2915 t^2 + 5108 t u
        + 297 u^2) + 12 s^6 (305 t^3\nonumber\\
  &   & + 873 t^2 u - 409 t u^2 - 105 u^3)
        + s^5 (5623 t^4 + 2463 t^3 u + 43455 t^2 u^2 - 13547 t u^3 + 
        6006 u^4)\nonumber\\
  &   & + 2 s^4 (9500 t^5 - 26299 t^4 u + 6042 t^3 u^2 + 14996 t^2 u^3 - 
        13858 t u^4 + 4347 u^5) + 6 s^3 (t + u)(1585 t^5\nonumber\\
  &   & - 3807 t^4 u - 2069
        t^3 u^2 + 3540 t^2 u^3 - 3635 t u^4 + 666 u^5)
        - 2 s^2 (t + u)^2 (3368 t^5\nonumber\\
  &   & - 6583 t^4 u + 1098 t^3 u^2 - 
        154 t^2 u^3 + 818 t u^4 - 291 u^5)+ 
        s t (t + u)^3 (1016 t^4 - 4048 t^3 u\nonumber\\
  &   & + 7194 t^2 u^2 - 5674 t u^3 + 
        2345 u^4) + 6 t^2 u (t + u)^4 (88 t^2 - 136 t u + 55 u^2)],\nonumber\\
L & = & -128 Q^{16} (9 s^3 - 286 s^2 t + 234 s t^2 - 768 t^3) + 
        64 Q^{14} [302 s^4 - 
        2 s^3 (892 t - 201 u) - s^2 t (329 t\nonumber\\
  &   & + 653 u) - 6 s t^2 (336 t - 337 u) 
        - 18 t^3 (188 t - 57 u)] - 
        32 Q^{12} [2164 s^5 - 26 s^4 (61 t - 104 u)\nonumber\\
  &   & - 3 s^3 (1565 t^2 - 456 t u - 81 u^2)+ 
        6 s^2 t (304 t^2 + 917 t u + 140 u^2) - 
        72 s t^2 (151 t^2 - 200 t u\nonumber\\
  &   & + 60 u^2) + 
        18 t^3 (272 t^2 + 729 t u + 217 u^2)] + 
        16 Q^{10} [6378 s^6 + 4 s^5 (3266 t + 1281 u)\nonumber\\
  &   & - 6 s^4 (208 t^2 - 2373 t u + 671 u^2) + 
        6 s^3 (3044 t^3 + 965 t^2 u + 877 t u^2 - 438 u^3) - 
        s^2 t (4111 t^3\nonumber\\
  &   & - 34593 t^2 u + 18147 t u^2 - 4367 u^3)+ 
        18 s t^2 (444 t^3 + 475 t^2 u - 2109 t u^2 - 460 u^3)\nonumber\\
  &   & + 6 t^3 (t + u)(346 t^2 - 1507 t u - 1754 u^2)] - 
        8 Q^8 [7745 s^7 + 4 s^6 (10507 t - 726 u) + 
        2 s^5 (16919 t^2\nonumber\\
  &   & + 7391 t u - 10689 u^2) + 2 s^4 (27235 t^3 - 7197 t^2 u - 
        8715 t u^2 - 4997 u^3)+ 
        s^3 (35626 t^4\nonumber\\
  &   & + 25776 t^3 u - 17340 t^2 u^2 + 40 t u^3 + 627 u^4)
        + 2 s^2 t (5453 t^4 - 7960 t^3 u - 
        31227 t^2 u^2\nonumber\\
  &   & + 14432 t u^3 - 1234 u^4) + 
        6 s t^2 (t + u)(1551 t^3 - 2225 t^2 u + 3494 t u^2 + 4615 u^3)
        \nonumber\\
  &   & - 18 t^3 (t + u)^2 (52 t^2 + 277 t u - 522 u^2)] + 
        4 Q^6 [88 s^8 + 2 s^7 (23351 t - 10201 u) + 
        3 s^6 (25513 t^2\nonumber\\
  &   & - 6487 t u - 9136 u^2) + 
        4 s^5 (23401 t^3 - 21672 t^2 u - 16431 t u^2 + 1690 u^3) + 
        2 s^4 (47681 t^4\nonumber\\
  &   & - 16807 t^3 u - 15504 t^2 u^2 - 6196 t u^3 + 9008 u^4)
        + 6 s^3 (8257 t^5 - 3337 t^4 u - 4869 t^3 u^2 + 
        5617 t^2 u^3\nonumber\\
  &   & - 3265 t u^4 + 723 u^5)+ 
        3 s^2 t (t + u)(6034 t^4 - 10736 t^3 u + 10784 t^2 u^2 - 5524 t u^3 -
        3909 u^4)\nonumber\\
  &   & - 18 s t^2 (t + u)^2 
        (62 t^3 + 460 t^2 u - 1051 t u^2 + 1413 u^3) - 18 t^3 (t + u)^3 
        (134 t^2 - 312 t u + 187 u^2)]\nonumber\\
  &   & + 2 Q^4 [7656 s^9 - 4 s^8 (2741 t - 5403 u) 
        - 3 s^7 (21617 t^2 - 12420 t u - 197 u^2)- 
        6 s^6 (15638 t^3\nonumber\\
  &   & - 19355 t^2 u - 4408 t u^2 + 6329 u^3)
        - 2 s^5 (55853 t^4 - 72369 t^3 u - 33240 t^2 u^2 + 
        19817 t u^3\nonumber\\
  &   & + 13359 u^4) - 6 s^4 (14998 t^5 - 13515 t^4 u - 
        6616 t^3 u^2 + 6070 t^2 u^3 + 4212 t u^4 - 121 u^5)\nonumber\\
  &   & - 3 s^3 (t + u)(10778 t^5 - 24948 t^4 u + 5428 t^3 u^2 - 12752 t^2 u^3
        + 8563 t u^4 - 945 u^5)\nonumber\\
  &   & + 2 s^2 t (t + u)^2 (1109 t^4 - 856 t^3 u - 9375 t^2 u^2 + 6659 t u^3 - 
        7720 u^4)+ 
        18 s t^2 (t + u)^3 (196 t^3\nonumber\\
  &   & - 662 t^2 u + 909 t u^2 - 486 u^3)
        - 6 t^3 (t + u)^4 (88 t^2 - 136 t u + 55 u^2)] - 
        Q^2 s [4672 s^9\nonumber\\
  &   & + 8 s^8 (1867 t + 675 u) - 4 s^7 (2290 t^2 - 6581 t u 
        + 2793 u^2) - 
        8 s^6 (6407 t^3 - 5820 t^2 u + 2055 t u^2\nonumber\\
  &   & + 1442 u^3)
        - s^5 (72785 t^4 - 122829 t^3 u + 7539 t^2 u^2 + 
        47513 t u^3 - 17400 u^4) - 4 s^4 (16577 t^5\nonumber\\
  &   & - 32131 t^4 u - 
        2523 t^3 u^2 + 22739 t^2 u^3 + 4412 t u^4 - 6702 u^5)
        - 2 s^3 (t + u)(14446 t^5 - 40612 t^4 u\nonumber\\
  &   & + 33235 t^3 u^2 + 9898 t^2 u^3
        + 2753 t u^4 - 5414 u^5)- 
        4 s^2 (t + u)^2 (76 t^5 - 1283 t^4 u\nonumber\\
  &   & - 1347 t^3 u^2 + 1924 t^2 u^3 - 
        1754 t u^4 - 264 u^5) + 
        s t (t + u)^3 (2504 t^4 - 9568 t^3 u + 14628 t^2 u^2\nonumber\\
  &   & - 11872 t u^3 + 
        5843 u^4) + 12 t^2 u (t + u)^4 (88 t^2 - 136 t u + 55 u^2)]- 
        s^2 [8 s^9 - 8 s^8 (509 t - 79 u)\nonumber\\
  &   & - 6 s^7 (971 t^2 + 944 t u - 259 u^2) + 
        4 s^6 (1348 t^3 + 21 t^2 u + 798 t u^2 + 133 u^3) + 
        s^5 (14551 t^4\nonumber\\
  &   & - 2921 t^3 u + 22191 t^2 u^2 + 973 t u^3 - 602 u^4)
        + 3 s^4 (3143 t^5 - 9528 t^4 u + 1430 t^3 u^2 + 3848 t^2 u^3\nonumber\\
  &   & - 3933 t u^4 + 560 u^5)+ 
        2 s^3 (t + u)(1072 t^5 - 12622 t^4 u + 7303 t^3 u^2 + 70 t^2 u^3 
        - 6706 t u^4 + 1855 u^5)\nonumber\\
  &   & + 2 s^2 (t + u)^2 (64 t^5 
        - 748 t^4 u + 3199 t^3 u^2 + 1036 t^2 u^3 
        - 2971 t u^4 + 1138 u^5)\nonumber\\
  &   & + 3 s u (t + u)^3 (616 t^4 - 1080 t^3 u + 828 t^2 u^2 - 287 t u^3 + 
        150 u^4) + 3 t u^2 (t + u)^4 (88 t^2\nonumber\\
  &   & - 136 t u + 55 u^2)],\nonumber\\
A & = & -256 Q^{16} s (47 s^2 - 160 s t + 168 t^2) + 
        128 Q^{14} s [481 s^3 - s^2 (394 t - 223 u) - 
        s t (608 t + 329 u)\nonumber\\
  &   & - 297 t^2 u] 
        - 192 Q^{12} s [623 s^4 + 2 s^3 (255 t + 92 u) 
        - 2 s^2 (147 t^2 - 379 t u + 35 u^2) - 
        2 s t (641 t^2\nonumber\\
  &   & + 46 t u - 280 u^2) - 
        3 t^2 (922 t^2 + 512 t u + 121 u^2)] + 
        32 Q^{10} s [3206 s^5 + s^4 (8171 t - 1281 u)\nonumber\\
  &   & + 3 s^3 (1611 t^2 + 678 t u - 1441 u^2) + 
        s^2 (16361 t^3 + 22752 t^2 u + 2829 t u^2 - 1294 u^3)\nonumber\\
  &   & - s t (30490 t^3 - 16863 t^2 u - 11742 t u^2 - 623 u^3) + 
        3 t^2 (332 t^3 + 6279 t^2 u + 5109 t u^2 + 1322 u^3)]\nonumber\\
  &   & - 16 Q^8 s [1447 s^6 + 11 s^5 (1441 t - 547 u) + 
        s^4 (5425 t^2 - 10841 t u - 7104 u^2) + 
        s^3 (72799 t^3\nonumber\\
  &   & - 4863 t^2 u - 25425 t u^2 + 721 u^3) - 
        s^2 (5911 t^4 - 62125 t^3 u + 37611 t^2 u^2 
        + 19939 t u^3 - 1171 u^4)\nonumber\\
  &   & - s t (10766 t^4 - 37108 t^3 u + 19803 t^2 u^2 + 29429 t u^3 + 
        7112 u^4) + 
        27 t^2 (t + u)(164 t^3 + 56 t^2 u\nonumber\\
  &   & - 497 t u^2 - 182 u^3)] 
        - 24 Q^6 s [866 s^7 - 2 s^6 (2379 t - 1481 u) + 
        s^5 (1161 t^2 + 1727 t u - 142 u^2)\nonumber\\
  &   & - s^4 (22889 t^3 - 28777 t^2 u + 595 t u^2 + 5245 u^3) - 
        2 s^3 (10917 t^4 - 3350 t^3 u - 14283 t^2 u^2 + 5145 t u^3\nonumber\\
  &   & + 1840 u^4) + 
        s^2 (3411 t^5 - 8839 t^4 u + 33722 t^3 u^2 - 906 t^2 u^3 - 8671 t u^4 - 
        673 u^5) - s t (t + u)(2542 t^4\nonumber\\
  &   & + 2224 t^3 u - 5790 t^2 u^2 + 6874 t
        u^3 + 3031 u^4) - 3 t^2 (t + u)^2 
        (478 t^3 - 1050 t^2 u + 399 t u^2\nonumber\\
  &   & + 307 u^3)] + 
        4 Q^4 s [3388 s^8 - 4 s^7 (497 t - 1419 u) + 
        3 s^6 (3527 t^2 + 54 t u - 217 u^2) - 
        s^5 (17365 t^3\nonumber\\
  &   & - 38457 t^2 u + 23991 t u^2 - 1835 u^3)
        - s^4 (58828 t^4 - 103125 t^3 u + 42771 t^2 u^2 + 
        32773 t u^3\nonumber\\
  &   & - 14379 u^4) - 3 s^3 (8829 t^5 - 1689 t^4 u - 
        27392 t^3 u^2 + 30108 t^2 u^3 - 417 t u^4 - 4199 u^5) +\nonumber\\ 
  &   & s^2 (t + u)(388 t^5 + 2699 t^4 u + 10138 t^3 u^2 - 30338 t^2 u^3 
        + 9539 t u^4 + 2992 u^5) - 
        s t (t + u)^2 (6512 t^4\nonumber\\
  &   & - 21184 t^3 u + 16905 t^2 u^2 - 3787 t u^3 - 
        4648 u^4) + 3 t^2 (t + u)^3 
        (532 t^3 - 600 t^2 u - 96 t u^2 + 199 u^3)]\nonumber\\
  &   & - 2 Q^2 s^2 [1064 s^8 + 8 s^7 (419 t + 7 u) + 
        2 s^6 (4483 t^2 + 2524 t u - 2607 u^2) - 
        4 s^5 (1025 t^3 + 3783 t^2 u\nonumber\\
  &   & - 3645 t u^2 + 1589 u^3) - 
        s^4 (18649 t^4 - 14389 t^3 u + 71991 t^2 u^2 - 36119 t u^3 + 
        4220 u^4)\nonumber\\
  &   & - 4 s^3 (2465 t^5 - 12601 t^4 u - 5511 t^3 u^2 + 13253 t^2 u^3 - 
        7972 t u^4 + 1290 u^5) - 2 s^2 (t + u)(1988 t^5\nonumber\\
  &   & - 4358 t^4 u - 5875 t^3
        u^2 + 3644 t^2 u^3 - 5942 t u^4 + 2063 u^5) 
        + 4 s (t + u)^2 (136 t^5\nonumber\\
  &   & + 942 t^4 u - 
        4783 t^3 u^2 + 2041 t^2 u^3 + 81 t u^4 - 259 u^5)
        + t (t + u)^3 (344 t^4 - 1792 t^3 u + 1848 t^2 u^2\nonumber\\
  &   & + 620 t u^3 - 853 u^4)].
\end{eqnarray}

$\nu_ls\to l^-c\overline{c}\left[{}^1\!S_0^{(8)}\right]c$:
\begin{eqnarray}
F & = & \frac{g^{\prime 4} \alpha_s^2 |V_{cs}|^2}{96\pi M s^4 (Q^2 + m_W^2)^2 
        (2 Q^2 + s - 2 t + u)^2 (2 Q^2 - 2 s + t + u)^4},\nonumber\\ 
T & = & 32 Q^{10} (2 s^3 + 10 s^2 t - 6 s t^2 - 3 t^3) + 
        16 Q^8 [5 s^4-s^3 (15 t - 26 u) - s^2 t (44 t + 17 u)\nonumber\\
  &   & - 3 s t^2 (3 t + 8 u) - 9 t^3 (t + u)] 
        - 24 Q^6 [14 s^5 + s^4 (17 t + 16 u)+ 2 s^3 (2 t^2 + 27 t u + u^2)
        \nonumber\\
  &   & + s^2 t (2 t + 5 u)^2+ 12 s t^2 u (t + u)+3 t^3 (t + u)^2] 
        + 4 Q^4 [44 s^6 + 8 s^5 (17 t - 6 u)\nonumber\\
  &   & + 3 s^4 (49 t^2 + 20 t u - 49 u^2) + s^3 (89 t^3 + 12 t^2 u - 135 t u^2 
        - 82 u^3) + s^2 t (t + u)(40 t^2 + 35 t u\nonumber\\
  &   & - 59 u^2)+ 
        3 s t^2 (t + u)^2 (t - 8 u)- 3 t^3 (t + u)^3]+ 
        2 Q^2 s [8 s^6- 4 s^5 (27 t - 20 u)\nonumber\\ 
  &   & - 2 s^4 (43 t^2 - 44 t u - 15 u^2) - s^3 (61 t^3 - 48 t^2 u - 3 t u^2 + 
        94 u^3) - 4 s^2 (t + u)(3 t^3 - 17 t^2 u\nonumber\\
  &   & - 16 t u^2 + 13 u^3)
        -s t (t + u)^2 (2 t^2 - 14 t u + 11 u^2)- 6 t^2 u (t + u)^3],\nonumber\\
L & = & 16 Q^{10} (4 s^3 - s^2 t + 12 s t^2 + 6 t^3) - 
        16 Q^8 [10 s^4- s^3 (15 t + 8 u)+ s^2 t (17 t - 10 u)\nonumber\\
  &   & - 3 s t^2 (3 t + 8 u)- 9 t^3 (t + u)] + 
        12 Q^6 [8 s^5- s^4 (41 t + 20 u)-10 s^3 t (2 t + 3 u)\nonumber\\
  &   & + s^2 t (t^2 + 26 t u + 36 u^2)+ 24 s t^2 u (t + u) + 6 t^3 (t + u)^2]+ 
        4 Q^4 [8 s^6 + s^5 (85 t + 24 u)\nonumber\\
  &   & + 6 s^4 (2 t + u)(11 t - 2 u)+ 
        s^3 (5 t^3 - 42 t^2 u - 18 t u^2 + 8 u^3)- 
        s^2 t (t + u)(2 t^2 - 41 t u - 79 u^2)\nonumber\\
  &   & - 3 s t^2 (t + u)^2 (t - 8 u) 
        + 3 t^3 (t + u)^3] - 
        Q^2 s [32 s^6+ 4 s^5 (21 t - 4 u)+ 4 s^4 (61 t^2\nonumber\\ 
  &   & - 8 t u - 15 u^2)+ 
        s^3 (215 t^3 + 150 t^2 u - 69 t u^2 - 88 u^3)+ 2 s^2 (t + u)^2 
        (18 t^2 - 11 t u - 38 u^2)\nonumber\\
  &   & +s t (t + u)^2 (40 t^2 + 8 t u - 77 u^2) 
        - 12 t^2 u (t + u)^3]+ 
        3 s^2 [4 s^5 t+4 s^4 (3 t^2 - u^2)\nonumber\\
  &   & + s^3 t (21 t^2 - 6 t u - 23 u^2)+ 
        s^2 (t + u)^2 (13 t^2 - 29 t u + 12 u^2)
        - s u (t + u)^2 (6 t^2 + t u - 8 u^2)\nonumber\\
  &   & + t u^2 (t + u)^3],\nonumber\\ 
A & = & 32 Q^{10} s (s^2 + 11 s t - 9 t^2) + 
        16 Q^8 s [13 s^3 - s^2 (18 t - 31 u)- s t (25 t + 22 u)
        - 9 t^2 (2 t + u)]\nonumber\\ 
  &   & - 24 Q^6 s [22 s^4 + s^3 (19 t + 26 u)+ s^2 (17 t^2 + 42 t u + u^2) + 
        5 s t (t^2 + 8 t u + 4 u^2)+3 t^2 (t^2 - u^2)]\nonumber\\ 
  &   & + 4 Q^4 s [76 s^5+ 
        4 s^4 (41 t + 3 u) + 3 s^3 (57 t^2 - 8 t u - 53 u^2) + 
        s^2 (100 t^3 + 33 t^2 u - 234 t u^2 - 95 u^3)\nonumber\\ 
  &   & + s t (t + u)(17 t^2 - 47 t u - 10 u^2)+ 9 t^2 u (t + u)^2] - 
        2 Q^2 s^2 [8 s^5 + 4 s^4 (33 t - 10 u) + 2 s^3 (41 t^2\nonumber\\ 
  &   & - 76 t u - 21 u^2)+ 
        s^2 (47 t^3 - 126 t^2 u - 141 t u^2 + 68 u^3)+ 
        2 s (t + u)(6 t^3 - 11 t^2 u + 32 t u^2 + 31 u^3)\nonumber\\ 
  &   & + t (t + u)^2 (4 t^2 - 4 t u - 17 u^2)].
\end{eqnarray}

$\nu_ls\to l^-c\overline{c}\left[{}^3\!S_1^{(8)}\right]c$:
\begin{eqnarray}
F & = & \frac{g^{\prime 4} \alpha_s^2 |V_{cs}|^2}{432 \pi M^3 s^4 
        (Q^2 + m_W^2)^2 (2 Q^2 + s - 2 t + u)^2 (2 Q^2 - 2 s + t + u)^4 
        (-4 Q^2 + s + t + 4 u)^2},\nonumber\\
T & = & -256 Q^{16} (68 s^3 - 100 s^2 t + 81 s t^2 - 129 t^3) + 
        128 Q^{14} [754 s^4 - s^3 (262 t - 337 u)\nonumber\\ 
  &   & + s^2 t (1303 t - 323 u) - 
        3 s t^2 (243 t - 74 u)- 18 t^3 (8 t + 43 u)] - 
        192 Q^{12} [1058 s^5 + s^4 (201 t\nonumber\\ 
  &   & + 814 u) + s^3 (1686 t^2 - 1297 t u + 134 u^2) + 
        s^2 t (485 t^2 + 2350 t u + 476 u^2)- 
        9 s t^2 (23 t^2\nonumber\\ 
  &   & + 119 t u - 115 u^2)+ 
        3 t^3 (31 t^2 + 76 t u + 33 u^2)] + 
        32 Q^{10} [6323 s^6 + 26 s^5 (121 t + 309 u)\nonumber\\ 
  &   & + 3 s^4 (2281 t^2 - 5643 t u + 482 u^2) + 
        s^3 (5849 t^3 + 12807 t^2 u - 11175 t u^2 - 3397 u^3)\nonumber\\ 
  &   & - s^2 t (1594 t^3 + 1497 t^2 u - 21309 t u^2 + 1171 u^3)+ 
        18 s t^2 (2 t + u)(18 t^2 + 43 t u - 17 u^2)\nonumber\\ 
  &   & + 3 t^3 (t + u)(77 t^2 + 736 t u + 344 u^2)] - 
        16 Q^8 [5698 s^7 + 2 s^6 (2839 t + 7232 u)\nonumber\\ 
  &   & + s^5 (2665 t^2 - 30689 t u + 4128 u^2) + 
        s^4 (1819 t^3 - 4218 t^2 u - 46863 t u^2 - 5648 u^3)\nonumber\\ 
  &   & - 2 s^3 (3065 t^4 + 7486 t^3 u - 5454 t^2 u^2 + 17027 t u^3 + 2044 u^4)
        - 4 s^2 t (758 t^4 + 2174 t^3 u\nonumber\\ 
  &   & + 3540 t^2 u^2 + 2411 t u^3 + 800 u^4)+ 
        3 s t^2 (t + u)(270 t^3 + 2297 t^2 u - 1223 t u^2 - 1783 u^3)\nonumber\\
  &   & + 9 t^3 (t + u)^2 (4 t^2 - 174 t u + 101 u^2)] + 
        24 Q^6 [364 s^8 + s^7 (2043 t + 4031 u) + s^6 (126 t^2\nonumber\\ 
  &   & - 6647 t u + 3871 u^2)
        -4 s^5 (1090 t^3 + 3820 t^2 u + 6803 t u^2 - 28 u^3) - 
        s^4 (1161 t^4 + 5383 t^3 u\nonumber\\ 
  &   & - 4443 t^2 u^2 + 10389 t u^3 - 2834 u^4) - 
        s^3 (4647 t^5 + 15326 t^4 u + 18464 t^3 u^2 + 6726 t^2 u^3\nonumber\\ 
  &   & - 2320 t u^4 - 2197 u^5)+ s^2 t (t + u)(37 t^4 + 973 t^3 u - 7686 t^2
        u^2 - 5822 t u^3 + 505 u^4)\nonumber\\ 
  &   & + 9 s t^2 (t + u)^2 (41 t^3 - 128 t^2 u + 129 t u^2 - 98 u^3)
        - 3 t^3 (t + u)^3 (43 t^2 + 10 t u + 162 u^2)]\nonumber\\ 
  &   & + 4 Q^4 [1288 s^9 - 4 s^8 (827 t + 309 u)- 24 s^7 (28 t^2 + 78 t u 
        + 581 u^2)+ 2 s^6 (4735 t^3 + 16281 t^2 u\nonumber\\ 
  &   & + 20409 t u^2 - 9989 u^3) + 
        s^5 (9605 t^4 + 51879 t^3 u + 52413 t^2 u^2 + 48737 t u^3 
        - 8922 u^4)\nonumber\\ 
  &   & + 6 s^4 (581 t^5 - 513 t^4 u - 6977 t^3 u^2 - 
        11518 t^2 u^3 - 6696 t u^4 - 1421 u^5)+ 
        s^3 (t + u)(7366 t^5\nonumber\\
  &   & + 27743 t^4 u + 47203 t^3 u^2 - 10811 t^2 u^3
        -40831 t u^4 - 8162 u^5)\nonumber\\ 
  &   & -s^2 t (t + u)^2 (3773 t^4 + 1034 t^3 u + 4305 t^2 u^2 - 3214 t u^3 + 
        6752 u^4)+9 s t^2 (t + u)^3 (144 t^3\nonumber\\ 
  &   & + 28 t^2 u - 6 t u^2 - 571 u^3)
        - 3 t^3 (t + u)^4 (38 t^2 - 20 t u + 41 u^2)] - 
        2 Q^2 s [512 s^9 - 4 s^8 (113 t\nonumber\\ 
  &   & - 539 u) - 4 s^7 (124 t^2 + 1255 t u + 1047 u^2)+ 
        s^6 (1267 t^3 + 5985 t^2 u + 2457 t u^2 - 15461 u^3)\nonumber\\ 
  &   & + s^5 (4181 t^4 + 21109 t^3 u + 22725 t^2 u^2 + 7343 t u^3 - 12422 u^4)
        + 2 s^4 (1088 t^5 + 4721 t^4 u\nonumber\\ 
  &   & + 5460 t^3 u^2 + 
        2156 t^2 u^3 + 3338 t u^4 + 2577 u^5)- 2 s^3 (t + u)(203 t^5 + 4561 t^4
        u + 11810 t^3 u^2\nonumber\\
  &   & + 20702 t^2 u^3 + 3997 t u^4 - 8011 u^5)+ 
        s^2 (t + u)^2 (772 t^5 + 10902 t^4 u + 10004 t^3 u^2\nonumber\\ 
  &   & - 6440 t^2 u^3 - 9579 t u^4 + 8075 u^5)- 
        s t (t + u)^3 (28 t^4 + 2464 t^3 u + 1554 t^2 u^2 
        + 346 t u^3\nonumber\\ 
  &   & - 4901 u^4) + 6 t^2 u (t + u)^4 (38 t^2 - 20 t u + 41 u^2)],\nonumber\\
L & = & -128 Q^{16} (41 s^3 - 73 s^2 t - 162 s t^2 + 258 t^3) + 
        64 Q^{14} [321 s^4 + s^3 (53 t + 311 u) - s^2 t (722 t\nonumber\\ 
  &   & - 1243 u)
        + 6 s t^2 (243 t - 74 u) + 36 t^3 (8 t + 43 u)] - 
        32 Q^{12} [841 s^5 + 7 s^4 (388 t + 189 u)\nonumber\\ 
  &   & - s^3 (846 t^2 - 8636 t u + 991 u^2)+ 
        3 s^2 t (1331 t^2 + 49 t u + 853 u^2)+ 
        54 s t^2 (23 t^2 + 119 t u - 115 u^2)\nonumber\\ 
  &   & - 18 t^3 (31 t^2 + 76 t u + 33 u^2)] + 
        16 Q^{10} [245 s^6 + 14 s^5 (628 t - 129 u) - 
        18 s^4 (26 t^2 - 1213 t u\nonumber\\ 
  &   & + 531 u^2) + s^3 (5597 t^3 - 1041 t^2 u + 5769 t u^2 - 5131 u^3) + 
        s^2 t (4562 t^3 + 15765 t^2 u - 26229 t u^2\nonumber\\ 
  &   & + 5273 u^3)- 36 s t^2 (2 t + u)(18 t^2 + 43 t u - 17 u^2) - 
        6 t^3 (t + u)(77 t^2 + 736 t u + 344 u^2)]\nonumber\\ 
  &   & + 8 Q^8 [3348 s^7 - s^6 (9239 t - 19819 u) 
        + 2 s^5 (1279 t^2 - 9989 t u + 18843 u^2) - 
        6 s^4 (144 t^3\nonumber\\ 
  &   & - 3613 t^2 u - 382 t u^2 - 3980 u^3)- 
        s^3 (6371 t^4 + 11822 t^3 u - 75102 t^2 u^2 + 4946 t u^3 
        - 2395 u^4)\nonumber\\ 
  &   & + s^2 t (2060 t^4 + 10469 t^3 u + 6942 t^2 u^2 + 5018 t u^3 + 3137 u^4)
        + 6 s t^2 (t + u)(270 t^3\nonumber\\
  &   & + 2297 t^2 u - 1223 t u^2 - 1783 u^3) + 
        18 t^3 (t + u)^2 (4 t^2 - 174 t u + 101 u^2)] - 
        4 Q^6 [8312 s^8\nonumber\\ 
  &   & + s^7 (113 t + 45201 u)
        + s^6 (5844 t^2 + 8461 t u + 75577 u^2) + 
        4 s^5 (2702 t^3 + 14361 t^2 u\nonumber\\ 
  &   & + 3267 t u^2 + 9992 u^3) + 
        s^4 (4607 t^4 + 30858 t^3 u + 141126 t^2 u^2 - 5122 t u^3 
        - 7605 u^4)\nonumber\\ 
  &   & + s^3 (3960 t^5 + 33346 t^4 u + 54082 t^3 u^2 + 42090 t^2 u^3 + 
        7969 t u^4 - 8561 u^5) + 3 s^2 t (t + u)(1646 t^4\nonumber\\
  &   & + 9812 t^3 u 
        - 3941 t^2 u^2 - 5333 t u^3 + 4713 u^4) + 
        54 s t^2 (t + u)^2 (41 t^3 - 128 t^2 u\nonumber\\ 
  &   & + 129 t u^2 
        - 98 u^3)- 18 t^3 (t + u)^3 (43 t^2 + 10 t u + 162 u^2)] + 
        2 Q^4 [9520 s^9 + 2 s^8 (2231 t\nonumber\\ 
  &   & + 24279 u) 
        + 18 s^7 (335 t^2 + 1501 t u + 4382 u^2)+ 
        s^6 (20915 t^3 + 73623 t^2 u + 14013 t u^2\nonumber\\ 
  &   & + 25049 u^3)+ 
        s^5 (14459 t^4 + 66834 t^3 u + 140442 t^2 u^2 - 51646 t u^3 - 
        46257 u^4)+ 9 s^4 (424 t^5\nonumber\\ 
  &   & + 5105 t^4 u + 9228 t^3 u^2 + 
        4392 t^2 u^3 - 3902 t u^4 - 4419 u^5)+ 
        s^3 (t + u)(12028 t^5 + 49382 t^4 u\nonumber\\
  &   & + 2989 t^3 u^2 - 33221 t^2 u^3
        + 15791 t u^4 - 8303 u^5) + 
        s^2 t (t + u)^2 (5650 t^4\nonumber\\ 
  &   & + 1756 t^3 u + 13398 t^2 u^2 - 
        5237 t u^3 + 3391 u^4)- 
        18 s t^2 (t + u)^3 (144 t^3 + 28 t^2 u - 6 t u^2\nonumber\\ 
  &   & - 571 u^3)
        + 6 t^3 (t + u)^4 (38 t^2 - 20 t u + 41 u^2)] - 
        Q^2 s [5424 s^9 + 4 s^8 (307 t + 6715 u)\nonumber\\ 
  &   & + 4 s^7 (818 t^2 + 3179 t u + 10041 u^2)+ 
        3 s^6 (5167 t^3 + 15237 t^2 u + 13 t u^2 - 2217 u^3)\nonumber\\ 
  &   & + s^5 (8507 t^4 + 55445 t^3 u + 87069 t^2 u^2 - 68617 t u^3 - 64252 u^4)
        + 2 s^4 (2522 t^5 + 16451 t^4 u\nonumber\\ 
  &   & + 24828 t^3 u^2 + 
        1754 t^2 u^3 - 28708 t u^4 - 23019 u^5)+ 
        6 s^3 (t + u)(1591 t^5 + 5269 t^4 u - 1673 t^3 u^2\nonumber\\
  &   & - 9487 t^2 u^3 
        + 1194 t u^4 - 592 u^5)+ 
        s^2 (t + u)^2 (3976 t^5 + 11984 t^4 u + 7492 t^3 u^2\nonumber\\ 
  &   & - 27182 t^2 u^3 - 
        15277 t u^4 + 3613 u^5)- 
        s t (t + u)^3 (1000 t^4 - 1136 t^3 u - 4314 t^2 u^2 
        - 3362 t u^3\nonumber\\ 
  &   & + 11911 u^4)- 12 t^2 u (t + u)^4 (38 t^2 - 20 t u + 41 u^2)] + 
        s^2 [544 s^9 + 4 s^8 (43 t + 831 u) + 4 s^7 (3 t^2\nonumber\\ 
  &   & - 49 t u + 1112 u^2) + 
        s^6 (1751 t^3 + 4593 t^2 u - 4467 t u^2 - 5821 u^3) + 
        s^5 (1490 t^4 + 10845 t^3 u\nonumber\\ 
  &   & + 15927 t^2 u^2 - 13873 t u^3 - 17133 u^4)
        + s^4 (741 t^5 + 3982 t^4 u + 8470 t^3 u^2 + 4356 t^2 u^3 - 
        7895 t u^4\nonumber\\ 
  &   & - 7886 u^5)+ 
        2 s^3 (t + u)(590 t^5 + 1567 t^4 u - 4234 t^3 u^2 - 8668 t^2 u^3 
        + 3472 t u^4 + 4391 u^5)\nonumber\\ 
  &   & + s^2 (t + u)^2 (562 t^5 + 2464 t^4 u - 
        894 t^3 u^2 - 13070 t^2 u^3 + 589 t u^4 + 9567 u^5)+ 
        s u (t + u)^3 (524 t^4\nonumber\\ 
  &   & + 848 t^3 u - 1230 t^2 u^2 - 1630 t u^3 + 
        2543 u^4) + 3 t u^2 (t + u)^4 (38 t^2 - 20 t u + 41 u^2)],\nonumber\\
A & = & 256 Q^{16} s (72 s^2 + 59 s t + 249 t^2) - 
        128 Q^{14} s [678 s^3 - s^2 (32 t - 69 u) + s t (1333 t 
        - 983 u)\nonumber\\ 
  &   & - 9 t^2 (5 t - 76 u)] + 192 Q^{12} s [909 s^4 - s^3 (277 t - 594 u) + 
        s^2 (1018 t^2 - 2285 t u + 138 u^2)\nonumber\\ 
  &   & - s t (616 t^2 - 346 t u + 889 u^2) - 9 t^2 (50 t^2 + 111 t u + 13 u^2)]
        - 32 Q^{10} s [5967 s^5 + s^4 (347 t\nonumber\\ 
  &   & + 9459 u) + 
        6 s^3 (884 t^2 - 2756 t u + 927 u^2) - 
        3 s^2 (349 t^3 - 3067 t^2 u + 1579 t u^2 - 933 u^3)\nonumber\\ 
  &   & - s t (5674 t^3 + 12369 t^2 u - 2715 t u^2 + 337 u^3) - 
        3 t^2 (61 t^3 + 1299 t^2 u + 2334 t u^2 + 988 u^3)]\nonumber\\ 
  &   & + 16 Q^8 s [7302 s^6 + 2 s^5 (3653 t + 10740 u) + 
        s^4 (7738 t^2 - 14657 t u + 20043 u^2) + 
        6 s^3 (1062 t^3\nonumber\\ 
  &   & + 2918 t^2 u - 2711 t u^2 + 890 u^3) - 
        2 s^2 (1498 t^4 - 6786 t^3 u - 30543 t^2 u^2 - 9941 t u^3 
        - 1884 u^4)\nonumber\\ 
  &   & + s t (247 t^4 - 9788 t^3 u - 10128 t^2 u^2 + 11500 t u^3 + 9325 u^4) + 
        9 t^2 (t + u)(83 t^3 + 402 t^2 u + 231 t u^2\nonumber\\
  &   & + 263 u^3)] - 
        24 Q^6 s [1476 s^7 + s^6 (4237 t + 8073 u) + 
        s^5 (2864 t^2 + 1595 t u + 10167 u^2)\nonumber\\ 
  &   & + s^4 (4117 t^3 + 6737 t^2 u - 7049 t u^2 + 3303 u^3) + 
        2 s^3 (170 t^4 + 5100 t^3 u + 9969 t^2 u^2 - 3700 t u^3\nonumber\\ 
  &   & - 2043 u^4) + 
        s^2 (2531 t^5 + 8522 t^4 u + 20322 t^3 u^2 + 22976 t^2 u^3 + 
        3530 t u^4 - 3171 u^5)\nonumber\\ 
  &   & + s t (t + u)(463 t^4 + 1795 t^3 u - 2562 t^2 u^2 + 928 t u^3 
        - 443 u^4) + 9 t^2 (t + u)^2 
        (10 t^3\nonumber\\ 
  &   & + 101 t^2 u + 277 t u^2 + 24 u^3)] + 
        4 Q^4 s [792 s^8 + 4 s^7 (2045 t + 3123 u) + 
        12 s^6 (699 t^2\nonumber\\ 
  &   & + 1642 t u + 1953 u^2) + 
        6 s^5 (425 t^3 + 289 t^2 u - 3805 t u^2 + 927 u^3) + 
        s^4 (8161 t^4 + 36723 t^3 u\nonumber\\ 
  &   & + 38145 t^2 u^2 - 37163 t u^3 - 18666 u^4)
        + 3 s^3 (629 t^5 + 3358 t^4 u + 7056 t^3 u^2 - 736 t^2 u^3\nonumber\\ 
  &   & - 13541 t u^4 - 6318 u^5)+ 3 s^2 (t + u)(2236 t^5 + 8557 t^4 u 
        + 3573 t^3 u^2 - 1745 t^2 u^3 - 9575 t u^4\nonumber\\
  &   & - 2154 u^5)- 
        s t (t + u)^2 (1378 t^4 + 1228 t^3 u - 6141 t^2 u^2 + 12280 t u^3 + 
        5149 u^4)\nonumber\\ 
  &   & + 3 t^2 (t + u)^3 (104 t^3 + 186 t^2 u - 426 t u^2 - 211 u^3)] 
        + 2 Q^2 s^2 [96 s^8 - 4 s^7 (355 t + 303 u)\nonumber\\ 
  &   & - 4 s^6 (934 t^2 + 2785 t u + 2367 u^2) + 
        3 s^5 (171 t^3 - 623 t^2 u + 1265 t u^2 - 581 u^3) - 
        s^4 (569 t^4\nonumber\\ 
  &   & - 5535 t^3 u - 18687 t^2 u^2 - 40237 t u^3 - 19446 u^4)- 
        4 s^3 (890 t^5 + 5930 t^4 u + 9786 t^3 u^2\nonumber\\ 
  &   & + 2525 t^2 u^3 - 3877 t u^4 - 1008 u^5)- 
        6 s^2 (t + u)(79 t^5 - 261 t^4 u - 2086 t^3 u^2 - 1178 t^2
        u^3\nonumber\\ 
  &   & + 1437 t u^4 + 2975 u^5)- s (t + u)^2 (256 t^5 + 3316 t^4 u 
        + 2856 t^3 u^2 
        - 19172 t^2 u^3 - 1367 t u^4\nonumber\\ 
  &   & + 8853 u^5)- 
        t (t + u)^3 (40 t^4 - 248 t^3 u - 966 t^2 u^2 + 1294 t u^3 + 1081 
u^4)].
\end{eqnarray}

$\nu_ls\to l^-c\overline{c}\left[{}^3\!P_J^{(8)}\right]c$:
\begin{eqnarray}
F & = & \frac{g^{\prime 4} \alpha_s^2 |V_{cs}|^2}{24 \pi M^3 s^4 
        (Q^2 + m_W^2)^2 (2 Q^2 + s - 2 t + u)^4 (2 Q^2 - 2 s + t + u)^5},
        \nonumber\\
T & = & -768 Q^{16} (8 s^3 + 2 s^2 t - 16 s t^2 - 29 t^3) + 
        384 Q^{14} [97 s^4 + 11 s^3 (11 t + 2 u)+ 
        s^2 t (382 t + 177 u)\nonumber\\ 
  &   & - s t^2 (21 t - 184 u)- 6 t^3 (76 t + 11 u)] - 
        576 Q^{12} [114 s^5 + s^4 (184 t - 33 u)+ 
        s^3 (147 t^2 - 421 t u\nonumber\\ 
  &   & - 134 u^2)+ 
        s^2 t (1405 t^2 + 587 t u - 144 u^2)- 
        s t^2 (425 t^2 - 449 t u - 10 u^2) + 
        t^3 (242 t^2 + 596 t u\nonumber\\ 
  &   & + 207 u^2)] + 
        96 Q^{10} [389 s^6 + 5 s^5 (157 t - 132 u) - 
        2 s^4 (1916 t^2 + 2421 t u + 162 u^2)+ 
        2 s^3 (3282 t^3\nonumber\\ 
  &   & - 4698 t^2 u - 1407 t u^2 + 304 u^3) - 
        s^2 t (646 t^3 + 4893 t^2 u + 8511 t u^2 + 637 u^3)+ 
        s t^2 (1139 t^3\nonumber\\ 
  &   & + 966 t^2 u - 5034 t u^2 - 1774 u^3) - 
        4 t^3 (t + u)(8 t^2 + 355 t u + 311 u^2)] + 
        48 Q^8 [68 s^7\nonumber\\ 
  &   & - s^6 (136 t - 725 u) + 
        s^5 (10631 t^2 + 2109 t u - 1206 u^2)+ 2 s^4 (2248 t^3 + 5992 t^2 u - 
        6168 t u^2\nonumber\\ 
  &   & - 1325 u^3)+ 2 s^3 (4294 t^4 + 9814 t^3 u - 
        4647 t^2 u^2 - 7169 t u^3 - 434 u^4)+ 
        s^2 t (584 t^4 + 4035 t^3 u\nonumber\\ 
  &   & + 3342 t^2 u^2 - 
        11746 t u^3 - 3258 u^4)- 
        s t^2 (t + u)(471 t^3 - 1622 t^2 u + 3626 t u^2 + 3676 u^3)\nonumber\\ 
  &   & - 3 t^3 (t + u)^2 (29 t^2 - 126 t u + 367 u^2)] - 
        72 Q^6 [96 s^8 - 10 s^7 (2 t - 3 u) + s^6 (3900 t^2 
        - 1483 t u\nonumber\\ 
  &   & + u^2)+ s^5 (3451 t^3 - 3472 t^2 u - 3477 t u^2 + 1314 u^3) + 
        s^4 (6017 t^4 - 1198 t^3 u - 12008 t^2 u^2\nonumber\\ 
  &   & + 2188 t u^3 + 2157 u^4) + 
        s^3 (2495 t^5 - 1676 t^4 u - 7964 t^3 u^2 - 3420 t^2 u^3 + 5105 t u^4 + 
        910 u^5)\nonumber\\ 
  &   & + s^2 t (t + u)(580 t^4 + 1156 t^3 u - 2443 t^2 u^2 + 1061 t u^3 
        + 1443 u^4)- 
        s t^2 (t + u)^2 (98 t^3\nonumber\\ 
  &   & - 276 t^2 u + 453 t u^2 - 1036 u^3)+ 
        2 t^3 (t + u)^3 (88 t^2 - 137 t u + 75 u^2)] 
        + 12 Q^4 [208 s^9\nonumber\\ 
  &   & - 4 s^8 (476 t - 159 u)+ 
        4 s^7 (1261 t^2 - 1584 t u + 279 u^2)+ 
        s^6 (3844 t^3 - 11529 t^2 u + 2658 t u^2\nonumber\\ 
  &   & - 553 u^3) + 
        s^5 (7955 t^4 - 27135 t^3 u - 5523 t^2 u^2 + 16307 t u^3 - 5196 u^4) 
        + s^4 (18632 t^5 - 5775 t^4 u\nonumber\\ 
  &   & - 4692 t^3 u^2 + 
        24368 t^2 u^3 + 7272 t u^4 - 6201 u^5) + 
        s^3 (t + u)(965 t^5 - 7412 t^4 u - 4906 t^3 u^2\nonumber\\
  &   & + 15238 t^2 u^3 - 2215
        t u^4 - 2246 u^5)+ 
        s^2 t (t + u)^2 (13 t^4 + 256 t^3 u - 2313 t^2 u^2 + 2299 t u^3 - 
        2516 u^4)\nonumber\\ 
  &   &+ s t^2 (t + u)^3 
        (592 t^3 - 1782 t^2 u + 2091 t u^2 - 1142 u^3)- 
        t^3 (t + u)^4 (232 t^2 - 232 t u + 67 u^2)]\nonumber\\ 
  &   & - 6 Q^2 s [272 s^9 - 8 s^8 (284 t - 91 u) - 4 s^7 (148 t^2 + 933 t u 
        - 63 u^2)+2 s^6 (1202 t^3 + 1255 t^2 u\nonumber\\ 
  &   & + 48 t u^2 - 349 u^3) + 
        s^5 (1045 t^4 - 2987 t^3 u + 13929 t^2 u^2 - 881 t u^3 + 214 u^4)+ 
        2 s^4 (2582 t^5\nonumber\\ 
  &   & - 8601 t^4 u + 30 t^3 u^2 + 4778 t^2 u^3 - 3600 t u^4 + 
        1071 u^5) + 2 s^3 (t + u)(2155 t^5 - 2009 t^4 u - 307 t^3 u^2\nonumber\\
  &   & + 1842 t^2
        u^3 - 3922 t u^4 + 1027 u^5) 
        - 2 s^2 (t + u)^2 (1064 t^5 - 761 t^4 u - 724 t^3 u^2\nonumber\\ 
  &   & - 190 t^2 u^3 + 856 t u^4 - 310 u^5)+ 
        s t (t + u)^3 (320 t^4 - 1776 t^3 u + 1890 t^2 u^2 
        - 1066 t u^3\nonumber\\ 
  &   & + 555 u^4) + 2 t^2 u (t + u)^4 (232 t^2 - 232 t u + 67 u^2)],\nonumber\\
L & = & -128 Q^{16} (40 s^3 + 67 s^2 t + 96 s t^2 + 174 t^3) + 
        64 Q^{14} [266 s^4+ 10 s^3 (61 t - 6 u)+ s^2 t (2153 t\nonumber\\ 
  &   & + 347 u) + 6 s t^2 (21 t - 184 u)+ 
        36 t^3 (76 t + 11 u)] - 32 Q^{12} [384 s^5 + 2 s^4 (129 t 
        - 653 u)\nonumber\\ 
  &   & + 20 s^3 (396 t^2 - 137 t u - 62 u^2)+ 
        3 s^2 t (235 t^2 - 1168 t u + 107 u^2)+ 
        18 s t^2 (425 t^2 - 449 t u\nonumber\\ 
  &   & - 10 u^2)- 
        18 t^3 (242 t^2 + 596 t u + 207 u^2)] - 
        16 Q^{10} [1022 s^6 + 2 s^5 (4519 t + 2706 u)\nonumber\\ 
  &   & - 6 s^4 (149 t^2 - 2885 t u - 414 u^2) + 
        4 s^3 (497 t^3 + 4173 t^2 u + 249 t u^2 - 535 u^3)- 
        s^2 t (859 t^3\nonumber\\ 
  &   & - 24765 t^2 u + 3333 t u^2 - 2489 u^3)+ 
        6 s t^2 (1139 t^3 + 966 t^2 u - 5034 t u^2 - 1774 u^3)\nonumber\\
  &   & - 24 t^3 (t + u)(8 t^2 + 355 t u + 311 u^2)] + 
        8 Q^8 [3364 s^7 + s^6 (26777 t + 5430 u) + 
        2 s^5 (21659 t^2\nonumber\\
  &   & + 9239 t u - 4422 u^2) 
        + 2 s^4 (14975 t^3 + 13824 t^2 u - 
        13008 t u^2 - 6814 u^3) + 2 s^3 (17155 t^4\nonumber\\
  &   & + 14097 t^3 u - 4659 t^2 u^2 - 
        2969 t u^3 - 1116 u^4) + s^2 t (3214 t^4 - 12410 t^3 u - 
        43626 t^2 u^2 + 15952 t u^3\nonumber\\ 
  &   & + 2941 u^4) + 
        6 s t^2 (t + u)(471 t^3 - 1622 t^2 u + 3626 t u^2 + 3676 u^3)
        + 18 t^3 (t + u)^2 
        (29 t^2\nonumber\\ 
  &   & - 126 t u + 367 u^2)] - 
        4 Q^6 [1872 s^8 + 8 s^7 (2931 t - 898 u) + 
        s^6 (88047 t^2 - 24599 t u\nonumber\\ 
  &   & - 23750 u^2)+ 6 s^5 (12809 t^3 - 3958 t^2 u - 
        12601 t u^2 - 1538 u^3) + 2 s^4 (43800 t^4 - 15493 t^3 u\nonumber\\ 
  &   & - 32520 t^2 u^2 - 9636 t u^3 + 5567 u^4)+ 2 s^3 (20619 t^5 
        + 7796 t^4 u - 
        11308 t^3 u^2 - 24 t^2 u^3\nonumber\\ 
  &   & - 3997 t u^4 + 2834 u^5)+ 
        3 s^2 t (t + u)(2800 t^4 - 4748 t^3 u + 10520 t^2 u^2 - 5494 t u^3 -
        4779 u^4)\nonumber\\
  &   & + 18 s t^2 (t + u)^2 (98 t^3 - 276 t^2 u + 
        453 t u^2 - 1036 u^3)- 36 t^3 (t + u)^3 
        (88 t^2\nonumber\\ 
  &   & - 137 t u + 75 u^2)]- 
        2 Q^4 [2656 s^9 + 8 s^8 (857 t + 1698 u)- 6 s^7 (10701 t^2 - 9155 t u -
        1344 u^2)\nonumber\\ 
  &   & - s^6 (90673 t^3 - 58884 t^2 u - 29715 t u^2 + 25426 u^3) 
        - 2 s^5 (49901 t^4 - 71901 t^3 u - 41961 t^2 u^2\nonumber\\ 
  &   & + 28391 t u^3 + 16518 u^4) - 6 s^4 (15176 t^5 - 7421 t^4 u - 
        15765 t^3 u^2 + 452 t^2 u^3 + 7555 t u^4\nonumber\\ 
  &   & + 1587 u^5) - 
        2 s^3 (t + u)(11653 t^5 - 22240 t^4 u + 232 t^3 u^2 - 10754 t^2 u^3 +
        11156 t u^4 - 476 u^5)\nonumber\\
  &   & + s^2 t (t + u)^2 (262 t^4 - 2336 t^3 u - 1326 t^2 u^2 + 6178 t u^3 - 
        14453 u^4)\nonumber\\ 
  &   & +6 s t^2 (t + u)^3 (592 t^3 - 1782 t^2 u + 2091 t u^2 - 1142 u^3)
        - 6 t^3 (t + u)^4 (232 t^2 - 232 t u + 67 u^2)]\nonumber\\ 
  &   & + Q^2 s [2720 s^9 + 128 s^8 (155 t + 27 u)- 4 s^7 (889 t^2 - 9125 t u 
        + 2274 u^2) - 4 s^6 (12176 t^3\nonumber\\ 
  &   & - 6261 t^2 u + 3138 t u^2 + 3383 u^3) 
        - s^5 (66007 t^4 - 107787 t^3 u + 17517 t^2 u^2 + 
        50095 t u^3\nonumber\\ 
  &   & - 10752 u^4) - 4 s^4 (16169 t^5 - 30391 t^4 u - 
        13023 t^3 u^2 + 22817 t^2 u^3 + 3506 t u^4 - 6942 u^5)\nonumber\\ 
  &   & - 2 s^3 (t + u)(14858 t^5 - 31832 t^4 u + 17273 t^3 u^2 + 23882 t^2 u^3
        + 1273 t u^4 - 8164 u^5)\nonumber\\ 
  &   & -4 s^2 (t + u)^2 (260 t^5 - 1171 t^4 u - 2955 t^3 u^2 + 3776 t^2 u^3 - 
        1528 t u^4 - 753 u^5)- 
        s t (t + u)^3 (208 t^4\nonumber\\ 
  &   & + 5392 t^3 u - 10164 t^2 u^2 + 9544 t u^3 - 
        5213 u^4)+ 12 t^2 u (t + u)^4 (232 t^2 - 232 t u + 67 u^2)]\nonumber\\ 
  &   & -s^2 [96 s^9 + 8 s^8 (519 t - 22 u) + 
        4 s^7 (1395 t^2 + 1447 t u - 260 u^2) - 6 s^6 (625 t^3 
        + 358 t^2 u\nonumber\\ 
  &   & + 649 t u^2 + 
        92 u^3) - s^5 (12099 t^4 - 2111 t^3 u + 23601 t^2 u^2 + 2067 t u^3 - 
        608 u^4)- s^4 (9513 t^5\nonumber\\ 
  &   & - 24944 t^4 u + 1486 t^3 u^2 + 
        11472 t^2 u^3 - 12575 t u^4 + 1312 u^5)- 6 s^3 (t + u)(596 t^5 - 3874
        t^4 u\nonumber\\
  &   & + 1681 t^3 u^2 + 604 t^2 u^3 - 2541 t u^4 + 648 u^5)- 
        2 s^2 (t + u)^2 (444 t^5 - 1964 t^4 u\nonumber\\ 
  &   & + 3031 t^3 u^2 + 2304 t^2 u^3 - 
        3716 t u^4 + 1444 u^5) - s u (t + u)^3 (608 t^4 - 1576 t^3 u 
        + 1812 t^2 u^2\nonumber\\ 
  &   & - 1321 t u^3 + 704 u^4) - 3 t u^2 (t + u)^4 
        (232 t^2 - 232 t u + 67 u^2)],\nonumber\\
A & = & 256 Q^{16} s (49 s^2 - 11 s t + 243 t^2) - 
        128 Q^{14} s [425 s^3 + s^2 (61 t + 68 u) - 
        7 s t (34 t + 163 u)\nonumber\\ 
  &   & + 9 t^2 (135 t - 14 u)] 
        + 192 Q^{12} s [565 s^4 + s^3 (753 t + 317 u)- 
        s^2 (624 t^2 + 227 t u - 241 u^2)\nonumber\\ 
  &   & - s t (134 t^2 + 1447 t u - 643 u^2)- 
        3 t^2 (771 t^2 + 779 t u + 239 u^2)] - 
        32 Q^{10} s [3523 s^5\nonumber\\ 
  &   & + 5 s^4 (2243 t + 348 u) + 
        6 s^3 (265 t^2 + 507 t u - 451 u^2) + 
        2 s^2 (9368 t^3 + 9495 t^2 u + 5907 t u^2\nonumber\\ 
  &   & - 1123 u^3)- 
        s t (25922 t^3 - 6987 t^2 u - 19623 t u^2 - 4183 u^3)+ 
        9 t^2 (285 t^3 + 1840 t^2 u\nonumber\\ 
  &   & + 1670 t u^2 + 556 u^3)] + 
        16 Q^8 s [2564 s^6 + 19 s^5 (1168 t - 235 u) + 
        s^4 (10973 t^2 - 3769 t u\nonumber\\ 
  &   & - 13434 u^2) + 
        2 s^3 (32611 t^3 - 948 t^2 u - 13905 t u^2 - 4025 u^3) + 
        s^2 (3319 t^4 + 51224 t^3 u\nonumber\\ 
  &   & - 24192 t^2 u^2 - 33026 t u^3 - 1321 u^4)- 
        s t (2599 t^4 - 34853 t^3 u + 9564 t^2 u^2 + 32050 t u^3\nonumber\\ 
  &   & + 12817 u^4)+ 
        9 t^2 (t + u)(180 t^3 - 211 t^2 u - 1187 t u^2 - 499 u^3)] + 
        24 Q^6 s [880 s^7\nonumber\\ 
  &   & - 10 s^6 (594 t - 455 u) - 
        s^5 (5508 t^2 - 3169 t u - 3085 u^2) - 
        s^4 (20143 t^3 - 24746 t^2 u\nonumber\\ 
  &   & - 5389 t u^2 + 6032 u^3)- 
        s^3 (22347 t^4 - 9890 t^3 u - 27216 t^2 u^2 + 9660 t u^3 
        + 8167 u^4)\nonumber\\ 
  &   & - s^2 (9 t^5 + 3014 t^4 u - 33328 t^3 u^2 - 3876 t^2 u^3 + 11363 t u^4 +
        2720 u^5) - s t (t + u)(2780 t^4\nonumber\\
  &   & + 1886 t^3 u - 4491 t^2 u^2 + 4661 t
        u^3 + 3803 u^4) - 
        3 t^2 (t + u)^2 (384 t^3 - 664 t^2 u\nonumber\\ 
  &   & + 415 t u^2 + 194 u^3)]- 
        4 Q^4 s [4544 s^8 + 4 s^7 (56 t + 2091 u) - 
        12 s^6 (481 t^2 - 792 t u\nonumber\\ 
  &   & + 419 u^2)- 
        s^5 (18620 t^3 - 40275 t^2 u + 15366 t u^2 + 11117 u^3)- 
        s^4 (51008 t^4 - 114369 t^3 u\nonumber\\ 
  &   & + 16449 t^2 u^2 + 
        36149 t u^3 - 11325 u^4)- 3 s^3 (12373 t^5 + 1095 t^4 u - 
        14926 t^3 u^2 + 30592 t^2 u^3\nonumber\\ 
  &   & + 669 t u^4 - 7111 u^5) + 
        s^2 (t + u)(2432 t^5 + 20995 t^4 u + 7298 t^3 u^2 - 34810 t^2 u^3 +
        5578 t u^4\nonumber\\
  &   & + 7739 u^5)-s t (t + u)^2 (8896 t^4 - 16580 t^3 u 
        + 10257 t^2 u^2 - 2351 t u^3 - 
        3821 u^4)\nonumber\\ 
  &   & + 9 t^2 (t + u)^3 
        (240 t^3 - 248 t^2 u + 68 t u^2 + 25 u^3)] + 
        2 Q^2 s^2 [1072 s^8 + 8 s^7 (832 t - 49 u)\nonumber\\ 
  &   & + 4 s^6 (1444 t^2 + 2143 t u - 1509 u^2) - 
        2 s^5 (4922 t^3 + 7689 t^2 u - 2772 t u^2 + 2165 u^3)\nonumber\\ 
  &   & - s^4 (15173 t^4 - 19913 t^3 u + 60255 t^2 u^2 - 
        21787 t u^3 - 920 u^4) - 
        4 s^3 (2242 t^5 - 13469 t^4 u\nonumber\\ 
  &   & - 5631 t^3 u^2 + 10354 t^2 u^3 - 
        6857 t u^4 + 807 u^5) - 2 s^2 (t + u)(3892 t^5 - 2872 t^4 u + 775 t^3
        u^2\nonumber\\
  &   & + 2434 t^2 u^3 - 7345 t u^4 + 3106 u^5)
        + 2 s (t + u)^2 (1168 t^5 + 3888 t^4 u - 
        8764 t^3 u^2\nonumber\\ 
  &   & + 2866 t^2 u^3 + 1776 t u^4 - 1153 u^5)- 
        t (t + u)^3 (272 t^4 + 2048 t^3 u - 2472 t^2 u^2\nonumber\\ 
  &   & + 800 t u^3 + 269 u^4)].
\end{eqnarray}

\end{appendix}

\newpage
\begin{figure}[ht]
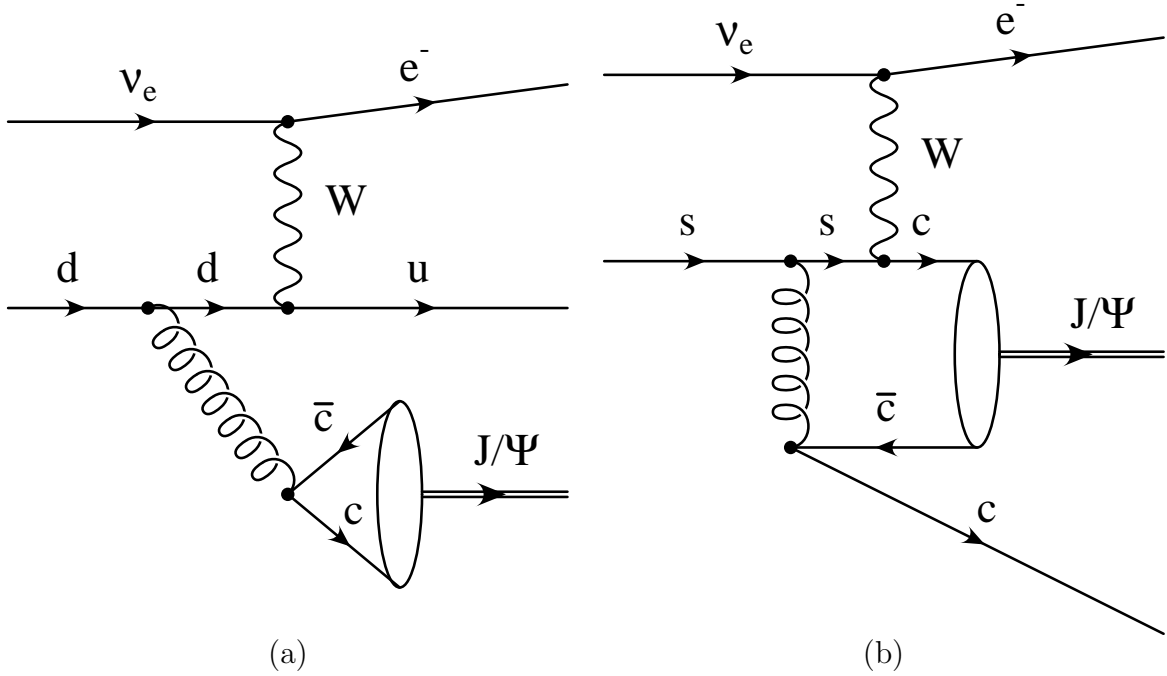

\begin{center}
\begin{tabular}{cc}
\parbox{7.5cm}{\epsfig{file=CC1.epsi,width=7.5cm}}
&
\parbox{7.5cm}{\epsfig{file=CC2.epsi,width=7.5cm}}
\\
(a) & (b)
\end{tabular}
\caption{Representative Feynman diagrams for the partonic subprocesses (a)
$\nu_e+d\to e^-+c\overline{c}[n]+u$ and (b)
$\nu_e+s\to e^-+c\overline{c}[n]+c$, where
$n={}^3\!S_1^{(1)},{}^3\!P_J^{(1)},{}^1\!S_0^{(8)},{}^3\!S_1^{(8)},
{}^3\!P_J^{(8)}$.
In case (a), there is one more Feynman diagram, which is obtained by
connecting the gluon line with the $u$-quark line.
In case (b), there are three more Feynman diagrams: two of type (a), where the
$d$ and $u$ quarks are replaced by the $s$ and $c$ quarks, respectively, and
one of type (b), where the gluon line is connected with the $c$-quark line to
the right of the $\overline{c}sW^+$ vertex.}
\label{fig:fey}
\end{center}
\end{figure}

\newpage
\begin{figure}[ht]
  \begin{center}
    \setlength{\unitlength}{1cm}
    \begin{picture}(15,10)
      \setlength{\unitlength}{1cm}
      \put(0.5,0){\includegraphics[width=15cm,height=10cm]{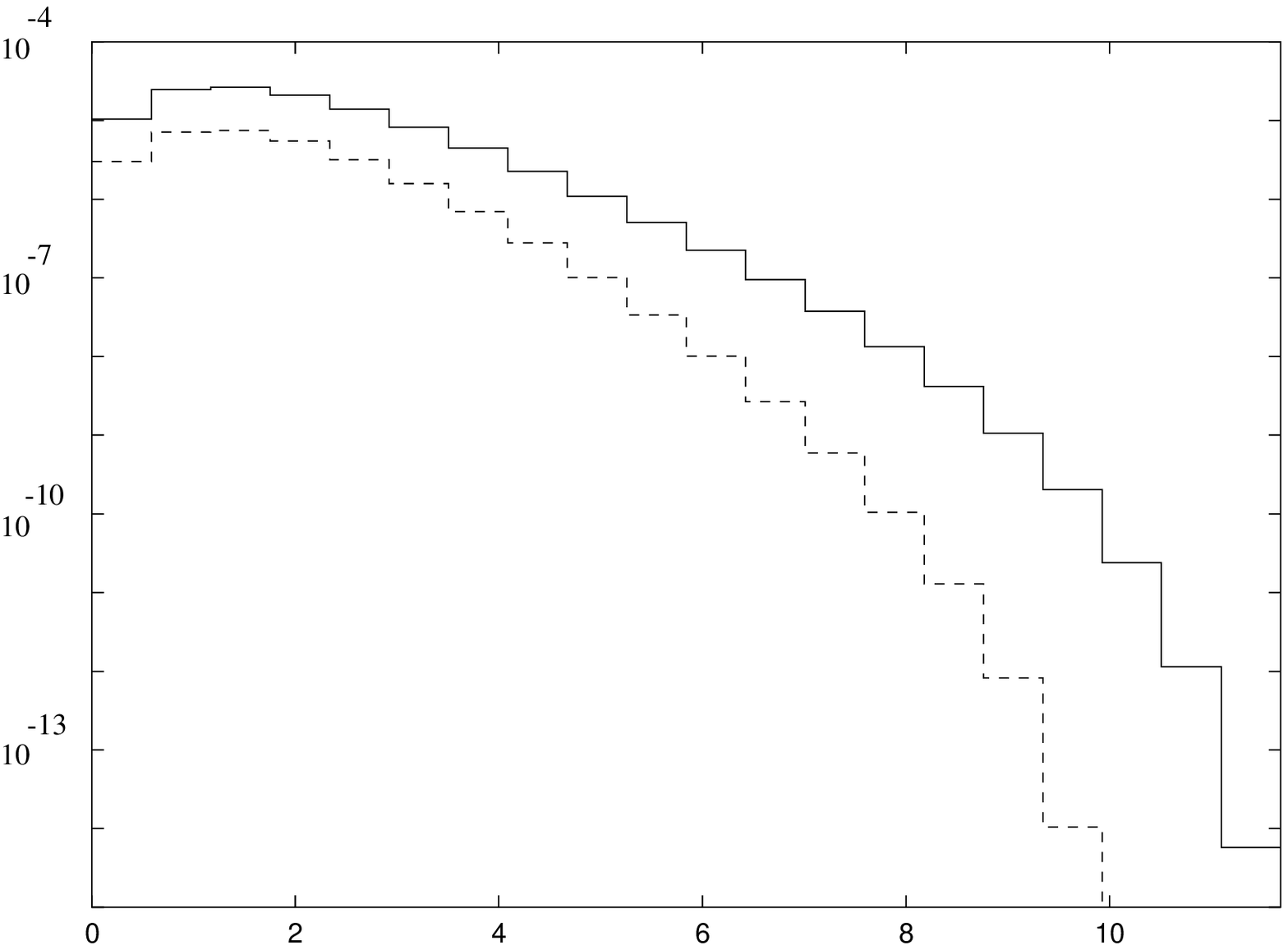}}
      \put(-1.5,5){$\displaystyle\frac{d\sigma}{dp_T}$}
      \put(-1.5,4){[fb/GeV]}
      \put(7.5,-0.5){$p_T$ [GeV]}
    \end{picture}
  \end{center}
\caption{NRQCD (upper histogram) and CSM (lower histogram) predictions for the
$p_T$ distribution of prompt $J/\psi$ inclusive production in $\nu N$ CC DIS
appropriate for the CHORUS experiment.}
\label{fig:nunpt}
\end{figure}

\newpage
\begin{figure}[ht]
  \begin{center}
    \setlength{\unitlength}{1cm}
    \begin{picture}(15,10)
      \setlength{\unitlength}{1cm}
      \put(0.5,0){\includegraphics[width=15cm,height=10cm]{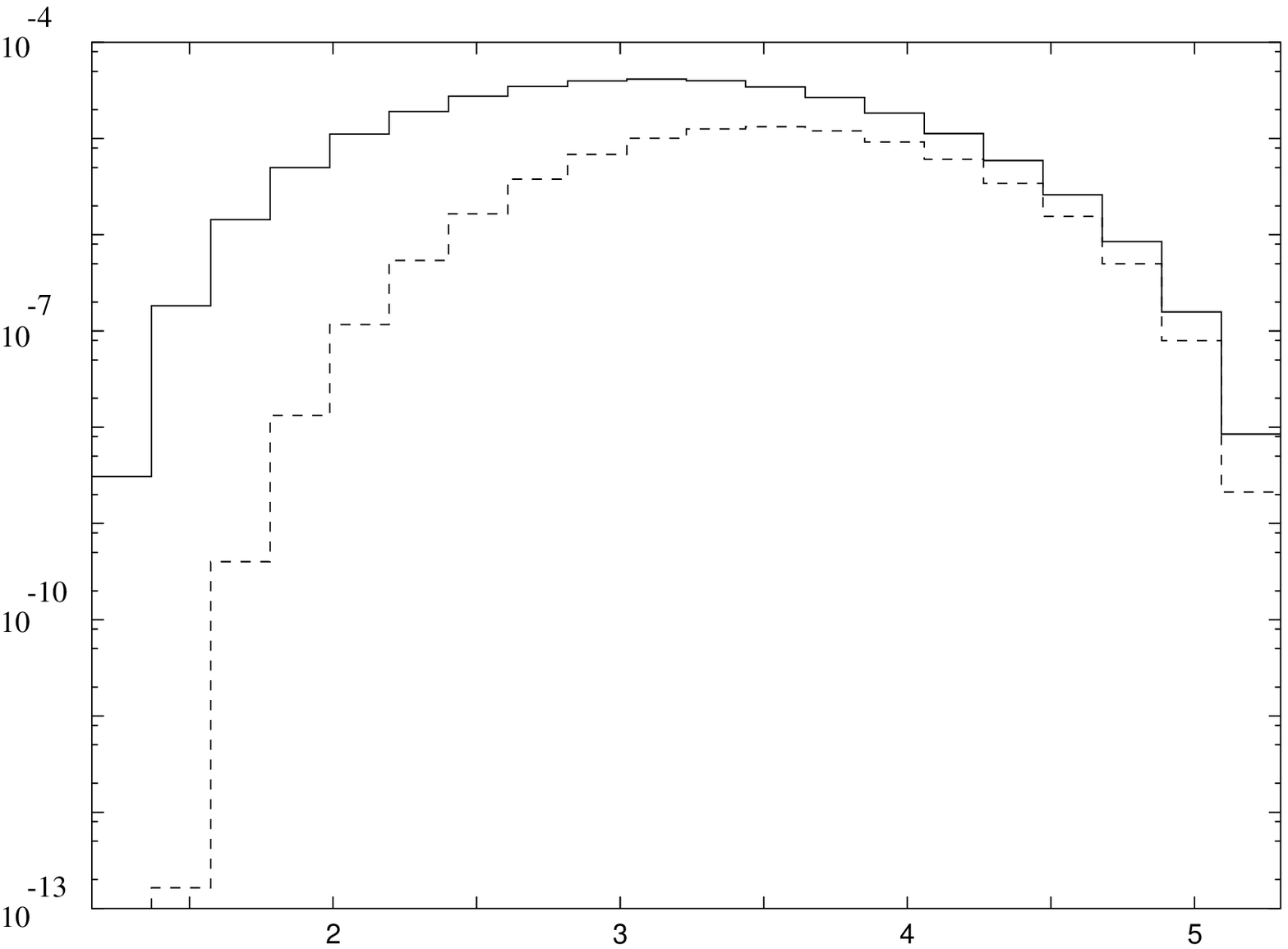}}
      \put(-0.5,5){$\displaystyle\frac{d\sigma}{dy}$}
      \put(-0.5,4){[fb]}
      \put(8.5,-0.5){$y$}
    \end{picture}
  \end{center}
\caption{NRQCD (upper histogram) and CSM (lower histogram) predictions for the
$y$ distribution of prompt $J/\psi$ inclusive production in $\nu N$ CC DIS 
appropriate for the CHORUS experiment.}
\label{fig:nuny}
\end{figure}

\newpage
\begin{figure}[ht]
  \begin{center}
    \setlength{\unitlength}{1cm}
    \begin{picture}(15,10)
      \setlength{\unitlength}{1cm}
      \put(0.5,0){\includegraphics[width=15cm,height=10cm]{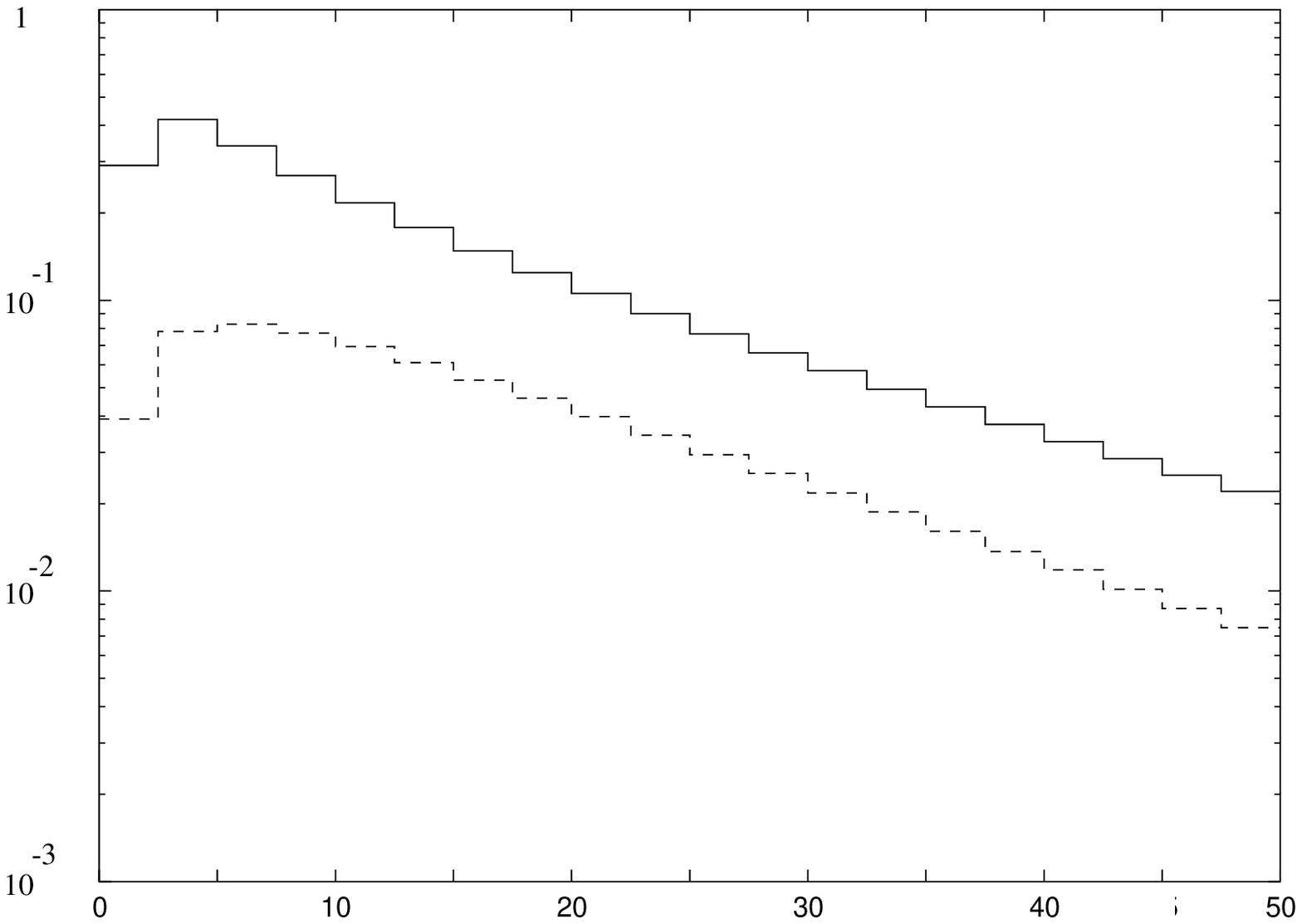}}
      \put(-1.5,5){$\displaystyle\frac{d\sigma}{dp_T}$}
      \put(-1.5,4){[fb/GeV]}
      \put(7.5,-0.5){$p_T$ [GeV]}
    \end{picture}
  \end{center}
\caption{NRQCD (upper histogram) and CSM (lower histogram) predictions for the
$p_T$ distribution of prompt $J/\psi$ inclusive production in $ep$ CC DIS
appropriate for the THERA experiment.}
\label{fig:eppt}
\end{figure}

\newpage
\begin{figure}[ht]
  \begin{center}
    \setlength{\unitlength}{1cm}
    \begin{picture}(15,10)
      \setlength{\unitlength}{1cm}
      \put(0.5,0){\includegraphics[width=15cm,height=10cm]{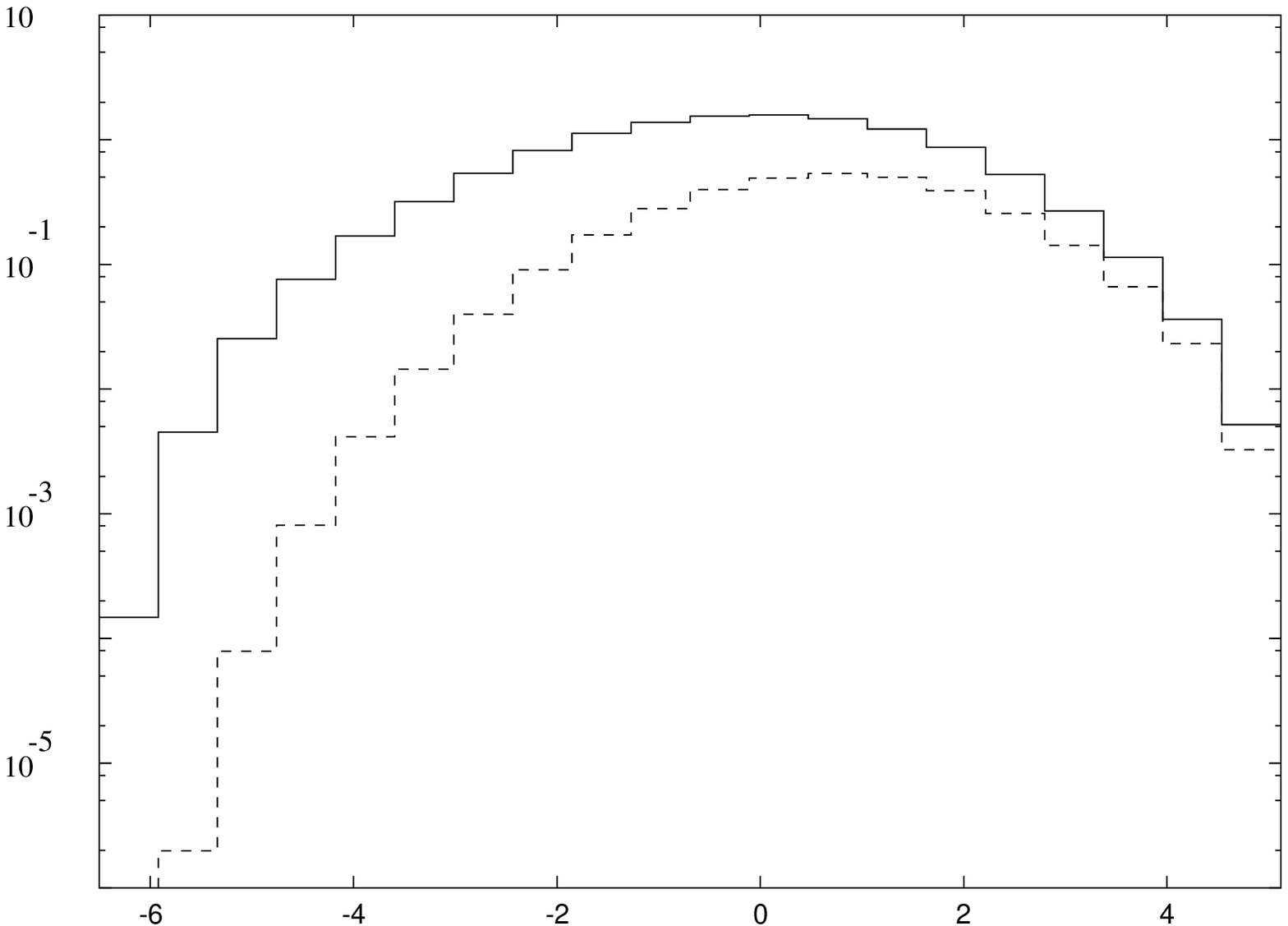}}
      \put(-0.5,5){$\displaystyle\frac{d\sigma}{dy}$}
      \put(-0.5,4){[fb]}
      \put(8.5,-0.5){$y$}
    \end{picture}
  \end{center}
\caption{NRQCD (upper histogram) and CSM (lower histogram) predictions for the
$y$ distribution of prompt $J/\psi$ inclusive production in $ep$ CC DIS
appropriate for the THERA experiment.}
\label{fig:epy}
\end{figure}

\end{document}